\documentclass[journal,10pt]{IEEEtran}

\usepackage[font=scriptsize,caption=false,labelsep=space]{subfig}
\usepackage{mathrsfs}
\usepackage{graphicx,float,wrapfig,epstopdf,amsmath}
\usepackage[]{algorithmicx}
\usepackage{algpseudocode,algorithm}
\usepackage{balance}

\usepackage{amsmath,mathtools }
\usepackage{amssymb}
\usepackage{amsthm}
\usepackage{graphicx}
\usepackage{epstopdf}
\usepackage{times}
\usepackage{textcomp,cite}
\usepackage{color}
\usepackage{url}
\usepackage{cuted}

\usepackage{multirow}
\usepackage[table]{xcolor}% http://ctan.org/pkg/xcolor
\usepackage{threeparttable}

\DeclareMathOperator*{\maximize}{maximize} % thin space, limits underneath in 
\DeclareMathOperator*{\minimize}{minimize} % thin space, limits underneath in 
\DeclareMathOperator*{\subjectto}{subject \hspace{3pt} to:} % thin space, limits
\begin{document}
	
	\title{A Deep Learning Framework for Hybrid Beamforming Without Instantaneous CSI Feedback 
		%in Multi-User mm-Wave Massive MIMO Systems
	}
	\author{Ahmet~M.~Elbir\textit{, Senior Member, IEEE}
		\thanks{A. M. E. is with the Department of Electrical and Electronics Engineering, Duzce University, Duzce, Turkey (e-mail: ahmetmelbir@gmail.com).}% 
	}
	% make the title area
	\maketitle
	
	\begin{abstract}
		Hybrid beamformer design plays very crucial role in the next generation millimeter-wave (mm-Wave) massive MIMO (multiple-input multiple-output) systems. Previous works assume the perfect channel state information (CSI) which results heavy feedback overhead. To lower complexity, channel statistics can be utilized such that only infrequent update of the channel information is needed. To reduce the complexity and provide robustness, in this work, we propose a deep learning (DL) framework to deal with both hybrid beamforming and channel estimation. For this purpose, we introduce three deep convolutional neural network (CNN) architectures. We assume that the base station (BS) has the channel statistics only and feeds the channel covariance matrix into a CNN to obtain the hybrid precoders. At the receiver, two CNNs are employed. The first one is used for channel estimation purposes and the another is employed to design the hybrid combiners. The proposed DL framework does not require the instantaneous feedback of the CSI {\color{black} at the BS. We have also investigated the online deployment of DL for channel estimation.} We have shown that the proposed approach has higher spectral efficiency with comparison to the conventional techniques. The trained CNN structures do not need to be re-trained due to the changes in the propagation environment such as the deviations in the number of received paths and the fluctuations in the received path angles up to 4 degrees. Also, the proposed DL framework exhibits at least 10 times lower computational complexity as compared to the conventional optimization-based approaches.

	\end{abstract}

	\section{Introduction}
	\label{sec:Introduciton}
	\IEEEPARstart{M}{illimeter wave} (mm-Wave) systems provide higher data rates, larger bandwidth and higher spectral efficiency as compared to the conventional cellular communications \cite{mimoOverview}. Hence, they become a promising candidate for the {\color{black} fifth generation} (5G) wireless communication systems~\cite{mimoOverview,5GwhatWillItBe,hodge2019reconfigurable}.  Compared to sub-6 GHz transmissions envisaged in 5G, the mm-Wave signals encounter a more complex propagation environment characterized by higher scattering, severe penetration losses, lower diffraction, and higher path loss for fixed transmitter and receiver gains \cite{mimoHybridLeus1,mimoHybridLeus2,mimoRHeath}. The mm-Wave systems leverage large-scale antenna arrays to compensate the propagation losses at high frequencies. However, the large number of antennas and high power consumption bring the difficulty of using a dedicated RF (radio frequency) chain for each antenna. In order to tackle this problem, hybrid (analog and baseband) beamforming architectures are introduced where small number of phase-only analog beamformers are used to steer the beams and process the down-converted signal via baseband beamformers, each of which is dedicated to a single RF chain \cite{mimoHybridLeus1,mimoHybridLeus2,mimoRHeath,mimoScalingUp}. 
	
	%	\subsection{Related Work}
	
	Hybrid beamforming is an effective approach to be used in mm-Wave systems, increasing the spectral efficiency and reduce the cost that could be imposed by large number of antennas in  massive multiple-input multiple-output (MIMO) systems \cite{mmwaveKeyElements,mimoRHeath}. In the literature, different approaches are proposed to design the hybrid beamformers in mm-Wave massive MIMO systems. One basic approach is selecting the columns of the RF precoder and combiners from a predefined codebook, which includes the array responses of receive/transmitted path angles.  \cite{mimoRHeath,mimoHybridLeus3,hybridBFLowRes}. However, the determination of the received path angles is very difficult in mm-Wave channel. To overcome this difficulty, phase extraction-based hybrid beamforming (PE-HB) techniques is proposed \cite{sohrabiNarrowband,hybridBFLowRes} without requiring such a codebook. In order to obtain an optimum solution, manifold optimization (MO) approach is proposed in \cite{hybridBFAltMin} where the Euclidean distance between the unconstrained beamformers and the hybrid beamformers (i.e.,	the multiplication of analog and baseband beamformer) is minimized. 
	
	%	In addition to the above works, deep learning (DL) based techniques are also proposed where a promising hybrid beamforming performance is obtained \cite{mimoDeepPrecoderDesign,mimoDLChannelModelBeamformingFacebook,elbirDL_COMML,elbirQuantizedCNN2019,elbirTVT}.
	Most of the above techniques assume that the instantaneous channel state information (CSI) is known a priori when designing the hybrid beamformers. Furthermore, the performance of these works strongly relies on the perfectness of the channel~\cite{channelEstLargeArrays,channelEstLargeArrays2,virtualPathSelection}. In practice, the pilot signals are periodically transmitted and the received data is processed to obtain the CSI~\cite{channelEstLargeArrays}. Hence, it is very crucial to perform  channel estimation accurately, especially in the presence of the challenges such as high data rate and short coherence intervals~\cite{coherenceTimeRef}. In order to cope with these challenges, statistical hybrid beamforming (SHB) architectures are proposed where the beamformers are designed by utilizing the channel statistics ~\cite{statisticalChannelModel2,statisticalChannelModel3,statisticalChannelModel1,widebandHBWithoutInsFeedback}. In this case, usually the second order statistics, i.e., the channel covariance matrices (CCMs) are used. Via CCM acquisition, the base station (BS) only knows the channel statistics, with infrequent channel information feedback, but no instantaneous CSI feedback. Hence, lower feedback overhead is achieved. In previous works, covariance-based beamforming	is considered, for instance in \cite{statisticalChannelModel2,statisticalChannelModel3}, baseband-only beamforming is proposed  where the receiver is assumed to have perfect CSI. Hybrid architectures with CCM is considered in \cite{statisticalChannelModel1,widebandHBWithoutInsFeedback}. In particular, \cite{statisticalChannelModel1} studies only the hybrid precoder design (without combiners)  and \cite{widebandHBWithoutInsFeedback} assumes the perfect CSI at the receiver, similar to \cite{statisticalChannelModel2,statisticalChannelModel3}. Furthermore, \cite{widebandHBWithoutInsFeedback} designs the analog precoders by simply taking the phases of the unconstrained precoders, which is a sub-optimum approach.
	
	%	\subsection{Motivation}
	In order to obtain low-complexity and effective hybrid beamforming performance, there is a need to design hybrid beamformers	without perfect CSI assumption. This motivates us to develop a hybrid beamforming algorithm using channel statistics at the BS  without the requirement of the perfect CSI at the receiver. To further obtain robust performance against the estimated/corrupted channel data, we design the hybrid beamformers via a deep learning (DL) approach.	DL has attracted many researchers in both communications and signal processing society due to its promising performance against many challenging problems such as channel estimation \cite{deepLearningChannelAndDOAEstHuang,channelEst_DL_superRes,mimoDLChannelEstimation,deepCNN_ChannelEstimation,mimoDLCSIFeedBack}, hybrid beamforming \cite{deepPrecoderDesignHuang,mimoDLChannelModelBeamformingFacebook,elbirDL_COMML,elbirQuantizedCNN2019,elbirTVT}. In particular, multilayer perceptrons (MLPs) have been proposed for hybrid precoding in \cite{deepPrecoderDesignHuang} and \cite{mimoDLChannelModelBeamformingFacebook}. The authors in  \cite{mimoDLChannelModelBeamformingFacebook} proposed a coordinated beam training approach via MLPs. In a recent work \cite{deepCNN_ChannelEstimation}, convolutional neural network (CNN) is designed  for channel estimation. In	\cite{elbirDL_COMML}, a CNN is designed for joint hybrid precoder and combiner design. A twin-CNN architecture is proposed in \cite{elbirQuantizedCNN2019} for joint antenna selection and hybrid beamforming, and multi-user hybrid beamforming is studied in \cite{elbirTVT} for mm-Wave massive MIMO systems. Note that the above DL-based beamforming approaches \cite{deepPrecoderDesignHuang,mimoDLChannelModelBeamformingFacebook} assume the perfect CSI to solve the hybrid beamformer design problem, even if this necessity is relaxed in \cite{elbirDL_COMML,elbirQuantizedCNN2019,elbirTVT} such that acceptable system rate performance can be achieved with corrupted/imperfect CSI via DL.	Thus, driven by the advantages of DL such as its provided low computational complexity and robustness against corrupted input data, we develop a DL framework for the hybrid beamformer design.
	
	%	\subsection{Contribution}
	In this paper, we introduce a DL framework where hybrid precoding/combining and channel estimation stages, which are very crucial tasks in mm-Wave communication systems, are performed via deep networks (Please see Fig.~\ref{fig_DLFramework}). We design three deep networks for this purpose. At the BS, a deep network called \textsf{CovNet} is used which accepts the input as the CCM and yields the hybrid precoders at the output. At the mobile station (MS), there are two deep networks, namely, \textsf{ChannelNet} and \textsf{BFNet}. \textsf{ChannelNet} is used in the channel training state to estimate the instantaneous CSI. The estimated channel matrix is then fed to \textsf{BFNet} to design the hybrid combiner weights at the output. As a result, the whole DL framework does not require either instantaneous CSI feedback or the perfect CSI at the receiver.

	The proposed DL framework has two stages: training (offline) and prediction (online). During training, several received pilot signals, channel and covariance realizations are generated, and hybrid beamforming problem is solved via manifold optimization (MO) approach \cite{hybridBFAltMin,manopt} to obtain the network labels. In the prediction stage, when the CNNs operate online, we estimate the hybrid beamformers and the channel matrix  by simply feeding the CNNs  with the related input data.	The proposed approach is advantageous since it does not require the perfect channel data in the prediction stage and still provides robust performance.		We summarize the main contributions of this paper as follows.
	\begin{enumerate}
		\item A DL framework is proposed which solves the hybrid beamformer design without instantaneous CSI feedback and does not require the perfect CSI at the receiver. Due to infrequent feedback of channel information, the proposed method has lower feedback overhead as compared to the conventional approaches~\cite{mimoRHeath,mimoHybridLeus3,hybridBFLowRes,sohrabiNarrowband,hybridBFAltMin,hybridBFLowRes}.
		
		\item The hybrid beamforming performance of the proposed method achieves higher spectral efficiency as compared to the state-of-the-art techniques such as both statistical~\cite{widebandHBWithoutInsFeedback} and non-statistical~\cite{sohrabiNarrowband} approaches.
		
		\item Unlike the other DL-based techniques~\cite{deepPrecoderDesignHuang,mimoDLChannelModelBeamformingFacebook,elbirDL_COMML,elbirQuantizedCNN2019,elbirTVT}, the proposed approach does not require the perfect knowledge of CSI. In fact, the proposed approach has a channel estimation stage taken place at the receiver.
		
		%		\item The proposed approach does not require a predefined codebook of array responses of received path angles as in the conventional techniques~\cite{mimoRHeath,mimoHybridLeus3}.
		%		
		
		\item The proposed DL approach provides more robust performance against the imperfections in the channel data as compared to the both DL-~\cite{elbirDL_COMML} and non-DL-based~\cite{sohrabiNarrowband} approaches. Together with superior performance, the proposed DL framework also enjoys less computation time.
		{\color{black}\item We have investigated the online deployment of the proposed DL-based channel estimation scheme where the deep network adapts itself to the propagation environment. }
		
	\end{enumerate}
	
	%	The rest of the paper is organized as follows. In the following section, we introduce the system model for mm-Wave channel and formulate the problem. Section~\ref{sec:bb_hb} presents the broadband hybrid beamformer design philosophy. We present our DL framework in Section~\ref{sec:HD_Design}. We follow this with numerical simulations in Section~\ref{sec:Sim} and conclude in Section~\ref{sec:Conc}.
	
	%		\subsection{Notation} 
	\textit{Notation:} Throughout the paper, vector and matrix quantities are denoted by boldface lower and upper case symbols, respectively. In the case of a vector $\mathbf{a}$, $[\mathbf{a}]_{i}$ represents its $i$-th element. For a matrix $\mathbf{A}$, $[\mathbf{A}]_{:,i}$ and $[\mathbf{A}]_{i,j}$ denote the $i$-th column and the $(i,j)$-th entry, respectively. {\color{black} $\mathbf{A}^*$, $\mathbf{A}^\textsf{T}$ and $\mathbf{A}^\textsf{H}$ represent the conjugate, transpose and Hermitian of $\mathbf{A}$. The Kronecker product is denoted by $\otimes$ while the Hadamard product is given by $\odot$.} $\mathbf{I}_N$ is the identity matrix of size $N\times N$, $\mathbb{E}\{\cdot\}$ denotes the statistical expectation, and $\|\cdot\|_\mathcal{F}$ is the Frobenious norm.  Finally, the notation $(\cdot)^{\dagger}$ denotes the Moore-Penrose pseudo-inverse while  $\angle\{\cdot\}$ denotes the angle of a complex scalar/vector. 
	%  the notation expressing  a convolutional layer with $N$ filters/channels of size $D\times D$, is given by  $N$@$ D\times D$.
	%		 For a complex scalar $a=e^{j\varphi}$ with continuous phase $\varphi$, $Q({a})= e^{j\varphi_B}$ denotes the quantization operator where $\varphi_B$ is the quantized angle in $[0,2\pi]$ sampled with $2^{B}$ points.
	
	%	
	%	
	%%-----------------------------------------------------	
	\begin{figure}[t!]
		\centering
		{\includegraphics[draft=false,scale=0.36]{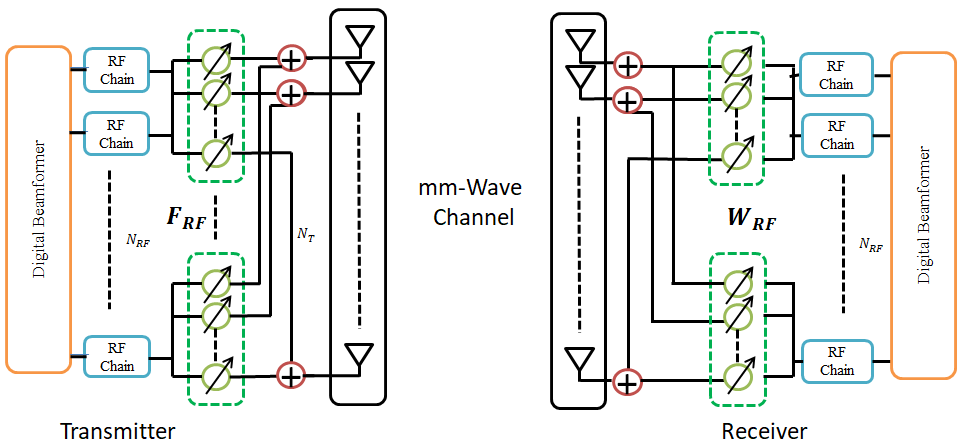} }
		\caption{System architecture of mm-Wave  MIMO based transceiver with hybrid  (analog and baseband) beamforming.}
		\label{fig_SystemArchitecture}
	\end{figure}
	%%-----------------------------------------------------	

	\section{System Model and Problem Formulation}
	\label{sec:SystemModel}
	We consider the hybrid beamformer design for {\color{black} a single-user} mm-Wave massive MIMO system as shown in Fig.~\ref{fig_SystemArchitecture}. The BS has $N_\mathrm{T}$ antennas and $N_\mathrm{RF}$ $(N_\mathrm{RF} \leq N_\mathrm{T})$ RF chains to transmit $N_\mathrm{S}$ data streams. In the downlink, the BS first precodes $N_\mathrm{S}$ data symbols $\mathbf{s} = [s_1,s_2,\dots,s_{N_\mathrm{S}}]^\textsf{T}\in \mathbb{C}^{N_\mathrm{S}}$ by applying the baseband precoder $\mathbf{F}_{\mathrm{BB}} = [\mathbf{f}_{\mathrm{BB}_1},\mathbf{f}_{\mathrm{BB}_2},\dots,\mathbf{f}_{\mathrm{BB}_{N_\mathrm{S}}} ]\in \mathbb{C}^{N_{\mathrm{RF}}\times N_\mathrm{S}}$. Then the baseband signal is conveyed via an RF precoder $\mathbf{F}_{\mathrm{RF}}\in \mathbb{C}^{N_\mathrm{T}\times N_{\mathrm{RF}}}$ to form the transmitted signal $	\mathbf{x} = \mathbf{F}_{\mathrm{RF}} \mathbf{F}_{\mathrm{BB}}  \mathbf{s}$. We assume that $\mathbf{F}_{\mathrm{RF}}$ consists of analog phase shifters, each of which has unit-modulus elements, i.e., $|[\mathbf{F}_{\mathrm{RF}}]_{i,j}|^2 =1$. Also, we have the power constraint  $||\mathbf{F}_{\mathrm{RF}}\mathbf{F}_{\mathrm{BB}} \|_\mathcal{F}^2= N_\mathrm{S}$  that is enforced by the normalization of the baseband precoder $\mathbf{F}_{\mathrm{BB}}$. 
	%	Thus, the $N_\mathrm{T}\times 1$ transmitted signal can be written as
	
	%		\footnote{We have used block-fading channel due to its common use in mm-Wave systems and  simplicity. We note here that the proposed DL approach can still work for different channel models since the main function of the deep network is mapping the input channel/covariance data to hybrid beamformer weights.}

	%		\subsection{Received Signal}
	Assuming a block-fading channel  model, the received signal at the MS is given by \cite{mmWaveModel1} 
	\begin{align}
	\label{arrayOutput}
	\bar{\mathbf{y} }= \sqrt{\rho}\mathbf{H} \mathbf{F}_\mathrm{RF}\mathbf{F}_\mathrm{BB}\mathbf{s} + \mathbf{n},
	\end{align}
	where $\rho$ represents the average received power, $\mathbf{H}\in \mathbb{C}^{N_\mathrm{R}\times N_\mathrm{T}}$ is the mm-Wave channel matrix and $\mathbf{n}\sim \mathcal{CN}(\mathbf{0},\sigma^2 \mathbf{I}_\mathrm{N_\mathrm{R}})$ is  additive white Gaussian noise (AWGN) vector.	At the receiver, the received signal is first processed by analog combiners $\mathbf{W}_\mathrm{RF}$, then the receiver employs low-dimensional $N_\mathrm{RF}\times N_\mathrm{S}$ digital combiners $\mathbf{W}_\mathrm{BB}$ to process the RF signal to obtain the received symbol vector as $\tilde{\mathbf{y}} = \mathbf{W}_\mathrm{BB}^\textsf{H}\mathbf{W}_\mathrm{RF}^\textsf{H}\bar{\mathbf{y}}$, i.e., 
	\begin{align}
	\label{sigModelReceived}
	\tilde{\mathbf{y}} =  \sqrt{\rho}\mathbf{W}_\mathrm{BB}^\textsf{H}\mathbf{W}_\mathrm{RF}^\textsf{H}\mathbf{H} \mathbf{F}_\mathrm{RF}\mathbf{F}_\mathrm{BB}\mathbf{s}  + \mathbf{W}_\mathrm{BB}^\textsf{H}\mathbf{W}_\mathrm{RF}^\textsf{H}\mathbf{n},
	\end{align}
	where the analog combiners $\mathbf{W}_\mathrm{RF}\in \mathbb{C}^{N_\mathrm{R}\times N_\mathrm{RF}}$ have element-wise constraint $\big[[\mathbf{W}_\mathrm{RF}]_{:,i}[\mathbf{W}_\mathrm{RF}]_{:,i}^\textsf{H}\big]_{i,i}=1$ similar to the RF precoders.

	\subsection{Channel Model}
	In mm-Wave transmission, the channel can be represented by the Saleh-Valenzuela (SV) model where a geometric channel model is adopted with limited scattering \cite{mimoChannelModel1,mimoChannelModel2}. Hence, we assume that the channel matrix  $\mathbf{H}$ includes the contributions of $L$ clusters, each of which has $N_\mathrm{sc} $ scattering paths/rays within the cluster. Thus, we can represent the downlink channel matrix by  an $N_\mathrm{R}\times N_\mathrm{T}$ matrix as
	\begin{align}
	\label{eq:ChannelModel}
	\mathbf{H} =  \sqrt{\frac{ N_\mathrm{T} N_{\mathrm{R}} } {N_{sc}L}}\sum_{l=1}^{L} \sum_{r=1}^{N_\mathrm{sc}}\alpha_{l,r}   \mathbf{a}_\mathrm{R}(\theta_{l,r}) \mathbf{a}_\mathrm{T}^\textsf{H}(\phi_{l,r}),
	\end{align} 
	where $\alpha_{l,r}\in \mathbb{C}$ denotes the complex gain corresponding to the $r$-th path in the $l$-th cluster, which are assumed to be independent zero-mean Gaussian random variables. $\mathbf{a}_\mathrm{R}(\theta)$ and $\mathbf{a}_\mathrm{T}(\phi)$ are the $N_\mathrm{R} \times 1$ and $N_\mathrm{T}\times 1$ steering vectors representing the array responses of the receive and transmit antenna arrays respectively. In particular, we define the steering vectors of receive and transmit arrays for a uniform linear array (ULA) as $\mathbf{a}_\mathrm{R}(\theta) = [1, e^{j\frac{2\pi}{\lambda} \bar{d}\sin(\theta)},\dots,e^{j\frac{2\pi}{\lambda} (N_\mathrm{R}-1)\bar{d}\sin(\theta)}]^\textsf{T}$ and $\mathbf{a}_\mathrm{T}(\phi) = [1, e^{j\frac{2\pi}{\lambda} \bar{d}\sin(\phi)},\dots,e^{j\frac{2\pi}{\lambda} (N_\mathrm{T}-1)\bar{d}\sin(\phi)}]^\textsf{T}$ respectively. Here, $\bar{d}$ is the uniform distance between the antennas and $\lambda = \frac{c_0}{f_c}$ is the wavelength for the carrier frequency $f_c$ with the speed of light $c_0$.

	\subsection{Problem Formulation}
	We formulate the main problem as designing the hybrid beamformers $\mathbf{F}_\mathrm{RF},\mathbf{F}_\mathrm{BB}$, $\mathbf{W}_\mathrm{RF},\mathbf{W}_\mathrm{BB}$ by maximizing the overall spectral efficiency of the system. which can be achieved by using the instantaneous channel matrix $\mathbf{H}$ available at the receiver. We assume that the Gaussian symbols are transmitted through the mm-Wave channel \cite{mimoRHeath,mimoHybridLeus1,mimoHybridLeus2}, thus, the hybrid beamformer design problem can be stated as follows
	\begin{align}
	\label{HBdesignProblem}
	&\underset{\mathbf{{F}}_\mathrm{RF},\mathbf{{W}}_\mathrm{RF}, \mathbf{{F}}_\mathrm{BB},\mathbf{{W}}_\mathrm{BB}}{\maximize}   \log_2 \bigg| \mathbf{I}_{N_\mathrm{S}} 
	+\frac{\rho}{N_\mathrm{S}}\boldsymbol{\Lambda}_\mathrm{n}^{-1}\mathbf{W}_\mathrm{BB}^\textsf{H}\mathbf{W}_\mathrm{RF}^\textsf{H}  \mathbf{H}\nonumber \\
	&\;\;\;\;\;\; \times\mathbf{{F}}_\mathrm{RF}\mathbf{{F}}_\mathrm{BB}\mathbf{{F}}_\mathrm{BB}^\textsf{H} \mathbf{{F}}_\mathrm{RF}^\textsf{H}\mathbf{H}^\textsf{H}\mathbf{W}_\mathrm{RF}\mathbf{W}_\mathrm{BB} \bigg|, \nonumber \\
	&\subjectto \; \mathbf{{F}}_\mathrm{RF} \in \mathcal{F}_\mathrm{RF},   \mathbf{{W}}_\mathrm{RF} \in \mathcal{W}_\mathrm{RF}, \nonumber \\
	&\;\;\;\;\;\;\;\;\;\;\;\;\;\;\;\;\;\;||\mathbf{{F}}_\mathrm{RF}\mathbf{{F}}_\mathrm{BB}||_{\mathcal{F}}^2 = N_\mathrm{S},
	\end{align}
	where $\boldsymbol{\Lambda}_\mathrm{n} = \sigma_n^2 \mathbf{W}_\mathrm{BB}^\textsf{H}\mathbf{W}_\mathrm{RF}^\textsf{H} \mathbf{W}_\mathrm{RF}\mathbf{W}_\mathrm{BB}\in \mathbb{C}^{N_\mathrm{S} \times N_\mathrm{S}}$ corresponds to the combiner-processed  noise term in the received signal (\ref{sigModelReceived}). $\mathcal{F}_\mathrm{RF}$ and $\mathcal{W}_\mathrm{RF}$ are the feasible sets for the RF precoder and combiners which obey the unit-modulus constraint. In some earlier works \cite{mimoRHeath,mimoHybridLeus3,elbirDL_COMML}, $\mathcal{F}_\mathrm{RF}$ and $\mathcal{W}_\mathrm{RF}$ are assumed to be known as the set of array responses of received/transmitted path angles. Then, the hybrid beamformers are designed by maximizing the spectral efficiency through a  greedy search over the columns of $\mathcal{F}_\mathrm{RF}$ and $\mathcal{W}_\mathrm{RF}$. In this paper, we do not have such an assumption. In fact, we design the hybrid beamformers via manifold optimization approach which does not require a predefined codebook as in \cite{mimoRHeath,mimoHybridLeus3,elbirDL_COMML}.
	
	In the proposed DL framework without instantaneous CSI feedback (Please see Fig.~\ref{fig_DLFramework}),  we make the following assumptions:
	
	\textbf{Assumption 1:} The BS and the MS do not have the knowledge of perfect CSI.
	
	\textbf{Assumption 2:} The BS {\color{black} does not require the CSI, instead it} only knows the spatial statistics of the channel, i.e., the covariance of the transmit antenna array at the BS is available through CCM acquisition\footnote{We assume that the CCM is available at the BS via CCM estimation approaches such as \cite{kalayci2019efficient}.}.
	
	\textbf{Assumption 3:} The MS can estimate the {\color{black} instantaneous channel} by processing the received pilot signals transmitted from the BS in the preamble stage.
	
	\textbf{Assumption 4:} We assume that the BS has the trained \textsf{CovNet} which accepts the CCM as input to design the hybrid precoders $\mathbf{F}_\mathrm{RF}$ and $\mathbf{F}_\mathrm{BB}$. The MS has the trained deep networks \textsf{ChannelNet} and \textsf{BFNet} to estimate the channel and design the hybrid combiners $\mathbf{W}_\mathrm{RF}$ and $\mathbf{W}_\mathrm{BB}$, respectively.
	
	In the following, we first discuss the channel estimation and hybrid beamforming, then we introduce our DL framework for the considered problem.

	\section{Channel Estimation and CCM Model}
	In practice, the estimation process of the channel matrix is a challenging task, especially in the case of a large number of antennas taking place in massive MIMO systems \cite{channelEstLargeArrays,channelEstimation1}. Furthermore, the coherence interval is very small in mm-Wave systems, making the channel estimation and acquisition process more difficult \cite{coherenceTimeRef}. In a practical scenario,  the estimated channel matrix can be obtained by channel estimation techniques \cite{mimoChannelModel2,channelEstimation1CS,channelEstimation1,angularDomainCE_feifei,mimoHybridLeus2}.
	
	\subsection{Channel Estimation}
	\label{sec:channelEstimation}
	In our DL framework, the channel estimation is performed by a deep network using the received pilot signals in the preamble stage. In this case, we assume the downlink scenario where the BS activates only one RF chain $\bar{\mathbf{f}}_u\in\mathbb{C}^{N_\mathrm{T}}$ to transmit pilot signals $\bar{{s}}_u$ on a single beam for $u = 1,\dots,M_\mathrm{T}$. Then, the receiver activates $M_\mathrm{R}$ RF chains to apply $\bar{\mathbf{w}}_v$ for $v = 1,\dots, M_\mathrm{R}$ to process the received pilots \cite{deepCNN_ChannelEstimation,mimoHybridLeus2,angularDomainCE_feifei}. Since the number of RF chains in the receiver is limited by $N_\mathrm{RF}$ ($< M_\mathrm{R}$), only $N_\mathrm{RF}$ combining vectors can be used at a single channel use. Hence, the total channel use in the channel acquisition process is $\lceil \frac{M_\mathrm{R}}{N_\mathrm{RF}}\rceil$. Then, the transmit and receive beamforming matrices become
	$\bar{\mathbf{F}} = [\bar{\mathbf{f}}_1,\bar{\mathbf{f}}_2,\dots,\bar{\mathbf{f}}_{M_\mathrm{T}}] \in \mathbb{C}^{N_\mathrm{T}\times M_\mathrm{T}}$ and $\bar{\mathbf{W}} = [\bar{\mathbf{w}}_1,\bar{\mathbf{w}}_2,\dots,\bar{\mathbf{w}}_{M_\mathrm{R}}]\in \mathbb{C}^{N_\mathrm{R}\times M_\mathrm{R}}$ respectively. Specifically, $\bar{\mathbf{F}}$ and $\bar{\mathbf{W}}$ can be constructed as the first $M_\mathrm{T}$ (or $M_\mathrm{R}$) column vectors of an $N_\mathrm{T}\times N_\mathrm{T}$ (or $N_\mathrm{R}\times N_\mathrm{R}$) discrete Fourier transform (DFT) matrix \cite{deepCNN_ChannelEstimation,angularDomainCE_feifei}.  Let us now consider the received signal in (\ref{sigModelReceived})  in the preamble as
	\begin{align}
	\label{receivedSignalPilot}
	\mathbf{\bar{Y}} = \bar{\mathbf{W}}^\textsf{H} \mathbf{H} \bar{\mathbf{F}}\bar{\mathbf{S}} + \tilde{\mathbf{N}},
	\end{align}
	where $\bar{\mathbf{S}} = \mathrm{diag}\{ \bar{s}_1,\dots,\bar{s}_{M_\mathrm{T}}\}$ denotes the pilot signals and $\tilde{\mathbf{N}}= \bar{\mathbf{W}}^\textsf{H} \bar{\mathbf{N}}$ is the effective noise matrix where $\bar{\mathbf{N}}$ denotes the AWGN matrix which corrupts the pilot training data by SNR$_{\bar{\mathbf{N}}}$. Without loss of generality, we assume  $\bar{\mathbf{S}} = \mathbf{I}_{M_\mathrm{T}}$, then the received signal in (\ref{receivedSignalPilot}) becomes
	\begin{align}
	\label{receivedSignalPilotMod}
	\mathbf{\bar{Y}} = \bar{\mathbf{W}}^\textsf{H} \mathbf{H}\bar{\mathbf{F}} + \tilde{\mathbf{N}}.
	\end{align}
	By processing $	\mathbf{\bar{Y}}$, we can obtain the initial channel estimate (ICE) as
	\begin{align}
	\label{Y}
	\mathbf{Y} = \mathbf{T}_\mathrm{T} \bar{\mathbf{Y}}\mathbf{T}_\mathrm{R},
	\end{align}
	where $\mathbf{T}_\mathrm{T} = \left\{
	\begin{array}{ll}
	\mathbf{\bar{W}},& M_\mathrm{R} < N_\mathrm{R} \\ (\mathbf{\bar{W}}\mathbf{\bar{W}}^\textsf{H})^{-1}\mathbf{\bar{W}}, & M_\mathrm{R} \geq N_\mathrm{R}.
	\end{array} 
	\right.$ and $\mathbf{T}_\mathrm{R} = \left\{
	\begin{array}{ll}
	\mathbf{\bar{F}}^\textsf{H},& M_\mathrm{T} < N_\mathrm{T} \\ \mathbf{\bar{F}}^\textsf{H}(\mathbf{\bar{F}}\mathbf{\bar{F}}^\textsf{H})^{-1}, & M_\mathrm{T} \geq N_\mathrm{T}.
	\end{array} 
	\right.$. We call $\mathbf{Y}$ initial channel estimate (ICE) since it will be used further in the proposed DL framework to obtain better channel estimate. Likewise, once $	\mathbf{Y}$ is obtained at the receiver, it is fed to the pretrained network \textsf{ChannelNet} to improve the channel estimation performance\footnote{\textsf{ChannelNet} is trained by accepting the input as initial channel estimate (ICE), $\mathbf{Y}$, and maps the input data to the labels which is the true channel matrix $\mathbf{H}$. As a result, a better channel estimation performance can be achieved as demonstrated in Section~\ref{sec:Sim}.}. Then the improved channel estimate is inserted to \textsf{BFNet} to obtain the hybrid combiners. 
	%	In the following, we discuss the design of the hybrid beamformers which are the labels of \textsf{BFNet}.

	%	\subsection{Statistical Channel Model}
	\subsection{CCM Model}
	\label{ccmAcquisition}
	Statistical beamforming strategies provide infrequent update of the channel information through the channel statistics, but no instantaneous feedback, hence, reducing  the feedback overhead~\cite{statisticalChannelModel1,statisticalChannelModel2,statisticalChannelModel3,widebandHBWithoutInsFeedback}. In practice, the CCM can be estimated by several algorithms such as temporal averaging techniques which collect the single snapshot received signals \cite{temporalAveragingCCM}, compressed covariance sensing approaches \cite{compressiveCCMSensing} and power angular spectrum estimation \cite{covarianceCCMChannelEst}, etc. Since the CCM acquisition is a certain field of research, in this paper we assume that the CCM is available at the BS, which can be obtained through above algorithms~\cite{temporalAveragingCCM,compressiveCCMSensing,covarianceCCMChannelEst,kalayci2019efficient}. In this work, we first exploit the structure of the CCM for hybrid precoder design, which will be employed in the proposed DL framework.
	
	Let us consider the channel model in (\ref{eq:ChannelModel}) which can be written as
	\begin{align}
	\mathbf{H} = \gamma \sum_{l = 1}^{L}\mathbf{A}_\mathrm{R}^{(l)} \boldsymbol{\Gamma}^{(l)} \mathbf{A}_\mathrm{T}^{{(l)}^\textsf{H}},
	\end{align}
	where $\gamma = \sqrt{\frac{ N_\mathrm{T} N_{\mathrm{R}} } {N_{sc}L}}$ and  $\mathbf{A}_\mathrm{R}^{(l)} = [\mathbf{a}_\mathrm{R}(\theta_{l,1}),\dots, \mathbf{a}_\mathrm{R}(\theta_{l,N_\mathrm{sc}})]$ and $\mathbf{A}_\mathrm{T}^{(l)} = [\mathbf{a}_\mathrm{T}(\phi_{l,1}),\dots, \mathbf{a}_\mathrm{T}(\theta_{l,N_\mathrm{sc}})]$ are $N_\mathrm{R}\times N_\mathrm{sc}$ and $N_\mathrm{T}\times N_\mathrm{sc}$ steering matrices of $N_\mathrm{sc}$ paths respectively. $\boldsymbol{\Gamma}\in \mathbb{C}^{N_{sc}\times N_{sc}}$ is a diagonal matrix which includes the path gains as $\boldsymbol{\Gamma}^{(l)} = \mathrm{diag} \{ \alpha_{l,1}, \dots, \alpha_{l,N_\mathrm{sc}}\}$. Using the property that the channel gains are independent random variables, we can write the covariance of the channel $\mathbf{R} = \mathbb{E}\{\mathbf{H}^\textsf{H} \mathbf{H}  \}$ as
	\begin{align}
	\label{Cov1}
	\mathbf{R} = \gamma^2 \sum_{l = 1}^{L} \mathbb{E}_\mathbf{H} \{ \mathbf{A}_\mathrm{T}^{(l)} \boldsymbol{\Gamma}^{(l)^\textsf{H}} \mathbf{A}_\mathrm{R}^{(l)^\textsf{H}} \mathbf{A}_\mathrm{R}^{(l)} \boldsymbol{\Gamma}^{(l)} \mathbf{A}_\mathrm{T}^{(l)^\textsf{H}}  \},
	\end{align}
	where the expectation is performed over $\mathbf{H}$. Incorporating the statistics of the AOA/AOD angles and the channel gains, we can rewrite (\ref{Cov1}) as 
	\begin{align}
	\mathbf{R} = \gamma^2\sum_{l = 1}^{L} \mathbb{E}_{\phi} \{ \mathbf{A}_\mathrm{T}^{(l)} \mathbb{E}_{\alpha}\{    \boldsymbol{\Gamma}^{(l)^\textsf{H}} \mathbb{E}_{\theta} \{\mathbf{A}_\mathrm{R}^{(l)^\textsf{H}} \mathbf{A}_\mathrm{R}^{(l)} \} \boldsymbol{\Gamma}^{(l)} \} \mathbf{A}_\mathrm{T}^{(l)^\textsf{H}}  \}. 
	\end{align}
	Since the receive steering vectors are unit-norm and normalized with $\frac{1}{\sqrt{N_\mathrm{R}}}$, we have $\mathbb{E}_{\theta} \{\mathbf{A}_\mathrm{R}^{(l)^\textsf{H}} \mathbf{A}_\mathrm{R}^{(l)} \} = \mathbf{I}_{N_\mathrm{sc}}$. Also, due to the independent zero-mean gains $\alpha$, we also have $\mathbb{E}_{\alpha}\{    \boldsymbol{\Gamma}^{(l)^\textsf{H}}  \boldsymbol{\Gamma}^{(l)} \} = \mathrm{diag}\{ \sigma_{\alpha_{l,1}}^2,\dots,\sigma_{\alpha_{l,N_\mathrm{sc}}}^2  \}$. Hence, we get
	\begin{align}
	\label{covariance}
	\mathbf{R} = \gamma^2 \sum_{l=1}^{L} \sigma_{\alpha_{l}}^2 \mathbb{E}\{\mathbf{A}_\mathrm{T}^{(l)} \mathbf{A}_\mathrm{T}^{(l)^\textsf{H}}  \}.
	\end{align}
	The CCM structure in (\ref{covariance}) explicitly implies that instantaneous channel information such as the path gains are not present. Hence, with comparison to the instantaneous channel in (\ref{eq:ChannelModel}), the covariance information in (\ref{covariance}) does not reflect the same precoding performance due to absence of the instantaneous channel gain information $\alpha_{l,r}$, instead the variance knowledge $\sigma_{\alpha_{l}}^2$. 
	
	{\color{black} Using only the covariance information in (\ref{covariance}), RF precoders $\mathbf{F}_\mathrm{RF}$ and $\mathbf{F}_\mathrm{BB}$ can be designed without requiring the instantaneous CSI feedback. This can be done by the infrequent feedback of $\sigma_{\alpha_{l}}^2$, mean AOD angles $\bar{\phi_{l}}$  and the angular spreads $\sigma_{\phi_{l}}$~\cite{widebandHBWithoutInsFeedback}. Then, the covariance matrix at the BS can be constructed as
		\begin{align}
		\mathbf{R} = \gamma^2 \sum_{l=1}^{L}\int_{\bar{\phi}_{l} - \frac{\sigma_{\phi_{l}}}{2}  }^{\bar{\phi}_{l} + \frac{\sigma_{\phi_{l}}}{2}}  \sigma_{\alpha_{l}}^2 \mathbf{A}_\mathrm{T}(\phi) \mathbf{A}_\mathrm{T}^\textsf{H}(\phi) d {\phi}.
		\end{align}
		To further reduce the infrequent feedback overhead, these statistical parameters can be quantized and the estimated CCM can be obtained as
		\begin{align}
		\hat{\mathbf{R}} = \gamma^2 \sum_{l=1}^{L}\sum_{\tilde{\phi}_l \in \mathcal{S}_{\phi_l} } \sigma_{\alpha_{l}}^2\mathbf{A}_\mathrm{T}(\tilde{\phi}) \mathbf{A}_\mathrm{T}^\textsf{H}(\tilde{\phi}),
		\end{align}
		where $\mathcal{S}_{\phi_l}$ is the set of discrete angles defined as
		\begin{align}
		\mathcal{S}_{\phi_l} = \{\tilde{\phi}_l: \bar{\phi_{l}} - \frac{\sigma_{\phi_{l}}}{2} , \bar{\phi}_{l} - \frac{\sigma_{\phi_{l}}}{2}  + {\delta}_{\phi_l}, \dots, \bar{\phi}_{l} + \frac{\sigma_{\phi_{l}}}{2}   \},
		\end{align}
		where ${\delta}_{\phi_l}$ is the angular resolution.
		
	}

	%	In our DL framework, we feed \textsf{CovNet} with $\mathbf{R}$  and obtain the hybrid precoders $\mathbf{F}_\mathrm{RF}$ and $\mathbf{F}_\mathrm{BB}$ at the output. 

	%	\subsection{Estimating The Channel Statistics}
	
	%	In our DL framework, the channel estimation is performed by a deep network which accepts the received pilot signals as input and yields the channel matrix estimate at the output layer~\cite{deepCNN_ChannelEstimation}. In the pilot transmission process, we assume that the transmitter only activates one RF chain to transmit the pilot on a single beam while all of the RF chains are activated in the receiver \cite{mimoHybridLeus2}. Hence, unlike the other DL-based works~\cite{elbirDL_COMML,elbirQuantizedCNN2019,mimoDLChannelModelBeamformingFacebook,mimoDeepPrecoderDesign} which assume the knowledge of the channel, in our framework, we jointly estimate both channel matrix as well as the hybrid beamformers by using deep networks.
	%	

	\section{Hybrid Beamformer Design For mm-Wave MIMO Systems}
	In this section, we first discuss the design of hybrid beamformers which will be, eventually, the labels of the proposed deep network architecture as discussed in Section~\ref{sec:Learning}. The design problem of the hybrid beamformers requires a joint optimization as in (\ref{HBdesignProblem}), however this approach is computationally prohibitive and even intractable. Instead, a decoupled problem is preferred \cite{mimoRHeath,sohrabiNarrowband,elbirQuantizedCNN2019,hybridBFAltMin}. Hence, in this work, we first design the hybrid precoders  $\mathbf{F}_\mathrm{RF},\mathbf{F}_\mathrm{BB}$ by utilizing the channel covariance matrix $\mathbf{R}$. Then, the receiver designs the hybrid combiners  $\mathbf{W}_\mathrm{RF},\mathbf{W}_\mathrm{BB}$ where  the channel matrix $\mathbf{H}$ is used. 
	
	\subsection{Hybrid Analog Precoder Design}
	In order to design the hybrid precoders, we rewrite the channel covariance matrix in (\ref{covariance}) as $\mathbf{R} = \gamma^2 \sum_{l=1}^{L} \sigma_{\alpha_{l}}^2 \mathbf{C}_l$ where $\mathbf{C}_l = \mathbb{E}\{ \mathbf{A}_\mathrm{T}^{(l)} \mathbf{A}_\mathrm{T}^{(l)^\textsf{H}}\}$. If the angular spread of the received paths is small, then $\mathbf{C}_l$ can be approximated as $\mathbf{C}_l\approx \mathbf{V}_l \boldsymbol{\Sigma}_l \mathbf{V}_l^\textsf{H}$, through eigendecomposition, corresponding to the largest few eigenvalues \cite{widebandHBWithoutInsFeedback}.  $\mathbf{V}_l$ denotes the eigenvectors of $\mathbf{C}_l$ corresponding to a few eigenvalues of $\mathbf{C}_l$ which are placed in descending order in $\boldsymbol{\Sigma}_l$. Then, we can write the approximate form of the channel covariance matrix $\mathbf{R}$ from (\ref{covariance}) as 
	\begin{align}
	\mathbf{R} \approx \gamma^2 \sum_{l=1}^{L}\sigma_{\alpha_{l}}^2 \mathbf{V}_l \boldsymbol{\Sigma}_l \mathbf{V}_l^\textsf{H} \approx \gamma^2 \mathbf{V} \boldsymbol{\Sigma} \mathbf{V}^\textsf{H},
	\end{align}
	where $\mathbf{V} = [\mathbf{V}_1,\dots, \mathbf{V}_L]$ and the diagonal elements of $\boldsymbol{\Sigma}$ are those of $\sigma_{\alpha_{l}}^2 \boldsymbol{\Sigma}_l$. Therefore, the optimum statistical beamformer $\mathbf{F}^\mathrm{opt}$ is the linear combination of the column vectors of $\mathbf{V}$, which can also be obtained from  the following problem
	\begin{align}
	\label{Fopt}
	\mathbf{F}^\mathrm{opt} = \arg \maximize_{\tilde{\mathbf{F}}} \hspace{2pt} ||\tilde{\mathbf{F}}^\textsf{H} \mathbf{R} \tilde{\mathbf{F}}||_\mathcal{F}^2 \nonumber
	\\
	\subjectto || \tilde{\mathbf{F}} ||_\mathcal{F}^2 = N_\mathrm{S}.
	\end{align}
	Once the unconstrained statistical beamformer is obtained, the next task is to determine the analog precoders $\mathbf{F}_\mathrm{RF}$. One possible solution is to solve 
	\begin{align}
	\label{Frf_suboptimum}
	\minimize_{\mathbf{F}_\mathrm{RF}} \hspace{3pt} || \mathbf{F}^\mathrm{opt} - \mathbf{F}_\mathrm{RF}  ||_\mathcal{F} \nonumber \\
	\subjectto  |[\mathbf{F}_\mathrm{RF}]_{i,j} |=1,
	\end{align}
	from which we can readily obtain the solution as $ \mathbf{F}_\mathrm{RF} = \angle \mathbf{F}^\mathrm{opt}$ which takes only the phase information of $\mathbf{F}^\mathrm{opt}$ \cite{widebandHBWithoutInsFeedback}. {\color{black} However, this approach is sub-optimum. In order to obtain an optimum solution for the RF precoder $\mathbf{F}_\mathrm{RF}$, we consider} the following problem, i.e.,
	\begin{align}
	\label{PrecoderDesign}
	&\underset{\mathbf{F}_\mathrm{RF},\mathbf{F}_\mathrm{BB}}{\minimize}\hspace{3pt} \big|\big|   \mathbf{F}^{\mathrm{opt}}  - \mathbf{F}_\mathrm{RF}\mathbf{F}_\mathrm{BB} \big|\big|_\mathcal{F}^2
	\nonumber \\
	&\subjectto \; \mathbf{{F}}_\mathrm{RF} \in \mathcal{F}_\mathrm{RF},  \nonumber \\
	&\big|\big| \mathbf{{F}}_\mathrm{RF}\mathbf{{F}}_\mathrm{BB}\big|\big|_{\mathcal{F}}^2 =  N_\mathrm{S}.
	\end{align}
	{\color{black} Above problem can be solved via alternating minimization approach where $\mathbf{F}_\mathrm{RF}$ and $\mathbf{F}_\mathrm{BB}$ are estimated one by one iteratively while one of them is fixed~\cite{hybridBFAltMin,manopt}. Although there are closed form expressions to estimate $\mathbf{F}_\mathrm{BB}$~\cite{mimoRHeath}, the estimation of $\mathbf{F}_\mathrm{RF}$ is  not straightforward due to the element-wise unit modulus constraint, i.e., $|[\mathbf{F}_\mathrm{RF}]_{i,j}|=1$. Hence, we first discuss the estimation of $\mathbf{F}_\mathrm{RF}$, by fixing $\mathbf{F}_\mathrm{BB}$. An effective solution can be achieved via conjugate gradient descent algorithm by using the tools of {\it Riemannian manifolds} \cite{optimizationManifolds,hybridBFAltMin,manopt}.   Let ${\bf x} = \mathrm{vec}\{\mathbf{F}_\mathrm{RF}\} \in \mathbb{C}^{\bar{M}}$ be the unknown vector to be optimized where $\bar{M} = N_\mathrm{T}N_\mathrm{RF}$, then the search space of (\ref{PrecoderDesign}) can be regarded as a Riemannian submanifold $\mathcal{M}$ of complex plane $\mathbb{C}^{\bar{M}}$ since ${\bf x}\in \mathbb{C}^{\bar{M}}$ forms a complex circle manifold, i.e., $\mathcal{M}_{cc}^{\bar{M}} = \{ {\bf x}\in \mathbb{C}^{\bar{M}} : |{\bf x}_1| = |{\bf x}_2|= \dots, |{\bf x}_{\bar{M}}| = 1\} $ \cite{optimizationManifolds}. The Riemannian gradient at $\bf x$, $\mathrm{grad}f({\bf x})$, can be defined as the orthogonal projection of the 
		Euclidean gradient $\nabla f({\bf x})$ onto the tangent space of ${\bf x}$, i.e.,
		\begin{align}
		\mathrm{grad}f({\bf x}) = \nabla f({\bf x}) - \operatorname{Re}\{ \nabla f({\bf x}) \odot {\bf x}^* \} \odot {\bf x},
		\end{align}
		where the Euclidean gradient of the cost function in (\ref{PrecoderDesign}) is given by
		\begin{align}
		\label{eq:gradF}
		\nabla f({\bf x}) = -2(\mathbf{F}_\mathrm{BB} \otimes \textbf{I}_{N_\mathrm{T}}) [ \mathbf{F}^{\mathrm{opt}} - (\mathbf{F}_\mathrm{BB}^\textsf{T} \otimes \textbf{I}_{N_\mathrm{T}}) {\bf x} ].
		\end{align}
		Then, the conjugate gradient descent algorithm \cite{optimizationManifolds} can be used where the unknown $\mathbf{x}$  is obtained with the update rule
		\begin{align}
		\label{Frf_update}
		{\bf x}_{l+1} = \frac{ ({\bf x}_l + \alpha_l {\bf d }_{l}) }  {|({\bf x}_l + \alpha_l {\bf d }_l)|},
		\end{align}
		where $\alpha_{l}$ is Armijo backtracking line search step size~\cite{manopt} and  $\mathbf{d}_l$ denotes the direction of decrease defined as
		\begin{align}
		{\bf d}_{l} =- \mathrm{grad}f({\bf x}_l) + \beta_l \bar{\bf d}_{l-1},
		\end{align}
		where $\mathrm{grad}f({\bf x}_l)$ denotes the Riemannian gradient at the $l$-th iteration and $\beta_l$ is the Polak-Ribiere parameter \cite{optimizationManifolds}. $\bar{\bf d}_{l}$ is the vector transport of conjugate direction ${\bf d}_l$ and it is defined as
		\begin{align}
		\bar{\bf d}_{l} = {\bf d}_{l} - \operatorname{Re}\{{\bf d}_{l} \odot  {\bf x}_{l+1}^*  \} \odot{\bf x}_{l+1},
		\end{align}
		where ${\bf x}_{l+1}$ can be directly obtained from (\ref{Frf_update}) and ${\bf d}_{0} =- \mathrm{grad}f({\bf x}_0)$. The optimization process can be initialized from a random point, i.e., $[\mathbf{x}_0]_{\bar{m}} = e^{j\bar{\theta}_{\bar{m}}}$ where $\bar{\theta}_{\bar{m}} \sim \mathrm{uniform}([0,2\pi))$, $\bar{m} = 1,\dots,\bar{M}$. 
		
		In (\ref{PrecoderDesign}), the convergence to an optimum solution is guaranteed such that the Euclidean distance between the unconstrained beamformer $\mathbf{F}^\mathrm{opt}$ and the hybrid beamformer $\mathbf{F}_\mathrm{RF}\mathbf{F}_\mathrm{BB}$ is minimized \cite{hybridBFAltMin}.

		\subsection{Hybrid Digital Precoder Design}
		Once $\mathbf{F}_\mathrm{RF}$ is obtained, the task is to estimate $\mathbf{F}_\mathrm{BB}$. While the optimum solution for $\mathbf{F}_\mathrm{BB}$ for instantaneous channel information is given in~\cite{fbb_optimum,alkhateeb2016frequencySelective}, here, we derive the solution by using the statistical channel information, i.e., $\mathbf{R}$.  Let us rewrite (\ref{Fopt}) by using $\mathbf{F}_\mathrm{RF}$ as
		\begin{align}
		\label{Fopt4Fbb}
		\mathbf{F}_\mathrm{BB} = \arg \maximize_{\mathbf{F}_\mathrm{BB}} \hspace{2pt} ||\mathbf{F}_\mathrm{BB}^\textsf{H}\mathbf{F}_\mathrm{RF}^\textsf{H} \mathbf{R} \mathbf{F}_\mathrm{RF} \mathbf{F}_\mathrm{BB}||_\mathcal{F}^2 \nonumber
		\\
		\subjectto ||\mathbf{F}_\mathrm{RF} \mathbf{F}_\mathrm{BB} ||_\mathcal{F}^2 = N_\mathrm{S},
		\end{align}
		which maximize the mutual information $\mathcal{I}(\mathbf{F}_\mathrm{RF}, \mathbf{F}_\mathrm{BB})$ at the BS, i.e.,
		\begin{align}
		\mathcal{I}(\mathbf{F}_\mathrm{RF}, \mathbf{F}_\mathrm{BB}) = \log_2 \bigg| \mathbf{I}_{N_\mathrm{S}} + 
		\frac{\rho}{N_\mathrm{S}}	  \mathbf{F}_\mathrm{BB}^\textsf{H}\mathbf{F}_\mathrm{RF}^\textsf{H} \mathbf{R} \mathbf{F}_\mathrm{RF} \mathbf{F}_\mathrm{BB}   \bigg|.
		\end{align}
		% (\ref{Fopt4Fbb}) is maximized if $\overline{\mathbf{D}} =\mathbf{F}_\mathrm{BB}^\textsf{H}\mathbf{F}_\mathrm{RF}^\textsf{H} \mathbf{R} \mathbf{F}_\mathrm{RF} \mathbf{F}_\mathrm{BB} $ is diagonal
		% 	Hence, an equivalent problem to (\ref{Fopt4Fbb}) can be written as
		%	\begin{align}
		%	\label{Fbb_SE}
		%	\mathbf{F}_\mathrm{BB}^\mathrm{opt} = \arg \maximize_{\mathbf{F}_\mathrm{BB}} \hspace{2pt} \mathcal{I}(\mathbf{F}_\mathrm{RF}, \mathbf{F}_\mathrm{BB}) \nonumber
		%	\\
		%	\subjectto ||\mathbf{F}_\mathrm{RF} \mathbf{F}_\mathrm{BB} ||_\mathcal{F} = N_\mathrm{S},
		%	\end{align}
		%	

		Let us decompose $\mathbf{F}_\mathrm{RF}$ as $		\mathbf{F}_\mathrm{RF} = \mathbf{U}_\mathrm{RF} \mathbf{Z}_\mathrm{RF},$
		%		\begin{align}
		%		\mathbf{F}_\mathrm{RF} = \mathbf{U}_\mathrm{RF} \mathbf{Z}_\mathrm{RF},
		%		\end{align}
		where $ \mathbf{U}_\mathrm{RF} =  \mathbf{F}_\mathrm{RF} (\mathbf{F}_\mathrm{RF}^\textsf{H}\mathbf{F}_\mathrm{RF})^{-\frac{1}{2}} $ and $\mathbf{Z}_\mathrm{RF} = (\mathbf{F}_\mathrm{RF}^\textsf{H}\mathbf{F}_\mathrm{RF})^{-\frac{1}{2}}$. Denoting $\mathbf{F}_\mathrm{BB} = (\mathbf{F}_\mathrm{RF}^\textsf{H}\mathbf{F}_\mathrm{RF})^{-\frac{1}{2}} \mathbf{G} $, we can write the following optimization problem
		\begin{align}
		\label{Fbb_SE}
		&\maximize_{\mathbf{G}} \hspace{2pt} \log_2 \bigg| \mathbf{I}_{N_\mathrm{S}} +  \frac{\rho}{N_\mathrm{S}}	\mathbf{G}^\textsf{H} \mathbf{U}_\mathrm{RF}^\textsf{H} \mathbf{R} \mathbf{U}_\mathrm{RF} \mathbf{G}\bigg|\nonumber
		\\
		&\subjectto ||\mathbf{U}_\mathrm{RF} \mathbf{G} ||_\mathcal{F}^2 = N_\mathrm{S}.
		\end{align}
		Let us define $N_\mathrm{S}\times N_\mathrm{S}$ matrix  $\mathbf{D}$ as $\mathbf{D}= \mathbf{G}^\textsf{H} \mathbf{M} \mathbf{G}$ where $\mathbf{M} = \mathbf{U}_\mathrm{RF}^\textsf{H} \mathbf{R} \mathbf{U}_\mathrm{RF}$. By Hadamard inequality~\cite{hadamard_inequality}, the optimum solution of (\ref{Fbb_SE}) can be only be achieved if $\mathbf{D}$ is diagonal. Also from \cite[Lemma 12]{fbb_optimum}, to form the diagonal matrix $\mathbf{D}$, for a given $\mathbf{G}$ and positive semidefinite Hermitian matrix $\mathbf{M}$, it is always possible to find another matrix $\tilde{\mathbf{G}}$ such that $\mathbf{G}^\textsf{H}\mathbf{M}\mathbf{G}= \tilde{\mathbf{G}}^\textsf{H}\mathbf{M}\tilde{\mathbf{G}}$ and $\mathrm{Trace}\{ \tilde{\mathbf{G}}\tilde{\mathbf{G}}^\textsf{H}\} \leq \mathrm{Trace}\{ {\mathbf{G}}{\mathbf{G}}^\textsf{H}\} $. Here, the optimum solution, i.e., $\mathbf{G} = \tilde{\mathbf{G}}$ can be found from the eigenvalue decomposition of $\mathbf{M}$ as $		\mathbf{G} = \mathbf{U}_\mathrm{M} \tilde{\boldsymbol{\Lambda}}_\mathrm{M},$
		%		\begin{align}
		%		\mathbf{G} = \mathbf{U}_\mathrm{M} \tilde{\boldsymbol{\Lambda}}_\mathrm{M},
		%		\end{align}
		for which we have $\mathbf{M} = \mathbf{U}_\mathrm{M} \boldsymbol{\Lambda}_\mathrm{M} \mathbf{U}_\mathrm{M}^\textsf{H}$ where  ${\mathbf{U}}_\mathrm{M}$ denotes the eigenvector matrix corresponding to the eigenvalues $\overline{\lambda}_1 \geq \overline{\lambda}_2 \geq \dots \geq  \overline{\lambda}_{N_\mathrm{S}}$  in $\boldsymbol{\Lambda}_\mathrm{M}= \mathrm{diag}\{\overline{\lambda}_1, \dots, \overline{\lambda}_{N_\mathrm{S}} \}$. $ \tilde{\boldsymbol{\Lambda}}_\mathrm{M} = \mathrm{diag}\{\tilde{\lambda}_1,\dots, \tilde{\lambda}_{N_\mathrm{S}}  \}$ is water-filling power allocation matrix \cite{fbb_optimum} with
		\begin{align}
		\tilde{\lambda}_n^2 = \max \bigg(0, \mu - \frac{N_\mathrm{S} }{\rho \bar{\lambda}_n^2}    \bigg), \hspace{3pt} n = 1,\dots,N_\mathrm{S},
		\end{align}
		where $\mu$ satisfies $\sum_{n=1}^{N_\mathrm{S}}\tilde{\lambda}_n^2 = ||\mathbf{F}_\mathrm{RF} \mathbf{F}_\mathrm{BB}||_\mathcal{F}^2 = N_\mathrm{S}$.
		Finally, we can select the optimum solution for $\mathbf{F}_\mathrm{BB}$ as 
		\begin{align}
		\label{Fbb}
		\mathbf{F}_\mathrm{BB} = (\mathbf{F}_\mathrm{RF}^\textsf{H}\mathbf{F}_\mathrm{RF})^{-\frac{1}{2}} \mathbf{U}_\mathrm{M} \tilde{\boldsymbol{\Lambda}}_\mathrm{M}.
		\end{align}

		By updating $\mathbf{F}_\mathrm{RF}$ and $\mathbf{F}_\mathrm{BB}$ from (\ref{Frf_update}) and (\ref{Fbb}), one can arrive an optimum solution in the sense that (\ref{PrecoderDesign}) is minimized~\cite{hybridBFAltMin,manopt}.

	}

	By designating the optimum solution $\mathbf{F}_\mathrm{RF}$ and $\mathbf{F}_\mathrm{BB}$ from (\ref{PrecoderDesign}) as the labels of the deep network, very good beamforming performance can be obtained~\cite{elbir2019lowcomplexity}.	Once the transmitter designs the hybrid precoders $\mathbf{F}_\mathrm{RF},\mathbf{F}_\mathrm{BB}$ via the channel covariance matrix, next we discuss the design of hybrid combiners at the receiver by using the estimated channel matrix  obtained from the preamble stage discussed in Section~\ref{sec:channelEstimation}.

	\subsection{Hybrid Combiner Design}
	At the receiver, the hybrid combiners are designed by utilizing only the estimated channel matrix from the received pilots. We can write the combiner design problem by minimizing the mean-square-error (MSE) between the transmitted and received-processed symbols $\{\mathbf{s}, \mathbf{W}_\mathrm{BB}^\textsf{H} \mathbf{W}_\mathrm{RF}^\textsf{H}\bar{\mathbf{y}}\}$ as follows
	\begin{align}
	\label{CombinerOnlyProblem}
	&\underset{\mathbf{W}_\mathrm{RF}, \mathbf{W}_\mathrm{BB} }{\minimize} \hspace{3pt}
	\mathbb{E}\{\big|\big| \mathbf{s} - \mathbf{W}_\mathrm{BB}^\textsf{H} \mathbf{W}_\mathrm{RF}^\textsf{H}\bar{\mathbf{y}}  \big|\big|_2^2\} \nonumber \\
	&\subjectto \mathbf{W}_\mathrm{RF} \in{\mathcal{W}}_\mathrm{RF}.
	\end{align}
	In order to write the problem in (\ref{CombinerOnlyProblem}) in more efficient form, we follow the steps in \cite{mimoRHeath} and add a constant term $\mathrm{Trace}\{\mathbf{W}_{\mathrm{MMSE}}^\textsf{H} \mathbb{E}\{\mathbf{y}\mathbf{y}^\textsf{H}\mathbf{W}_{\mathrm{MMSE}}  \} \} - \mathrm{Trace}\{\mathbf{s}\mathbf{s}^\textsf{H} \}$ into the cost function in (\ref{CombinerOnlyProblem}). Here, $\mathbf{W}_{\mathrm{MMSE}}$ denotes the MMSE estimator defined as $\mathbf{W}_\mathrm{MMSE}= (\mathbb{E}\{\mathbf{s} \bar{\mathbf{y}}^\textsf{H} \} \mathbb{E}\{\bar{\mathbf{y}}  \bar{\mathbf{y}}^\textsf{H}   \}^{-1})^\textsf{H}$ which can be written  in a compact form as,
	\begin{align}
	&\mathbf{W}_\mathrm{MMSE}^\textsf{H} = \nonumber \\
	& \frac{1}{\rho}\bigg( \mathbf{F}_\mathrm{BB}^{\textsf{H}}\mathbf{F}_\mathrm{RF}^\textsf{H}  \mathbf{H}^\textsf{H}\mathbf{H}\mathbf{F}_\mathrm{RF}\mathbf{F}_\mathrm{BB} +
	\frac{N_\mathrm{S}\sigma_n^2}{\rho}\mathbf{I}_{N_\mathrm{S}} \bigg)^{-1}  \mathbf{F}_\mathrm{BB}^{\textsf{H}}\mathbf{F}_\mathrm{RF}^\textsf{H} \mathbf{H}^\textsf{H}. \nonumber
	\end{align}
	
	%	By incorporating the unconstrained beamformer $\mathbf{F}^{\mathrm{opt}}$\footnote{In conventional works \cite{mimoRHeath,mimoHybridLeus3} where instantaneous feedback is assumed, instead of $\mathbf{F}^{\mathrm{opt}}$, the unconstrained beamformer from the singular value decomposition (SVD) of the channel matrix, is used. Since we assume no , The unconstrained beamformer $\mathbf{F}^{\mathrm{opt}}$} from (\ref{Fopt}), 
	
	Then, an equivalent problem to  (\ref{CombinerOnlyProblem}) can be stated as follows
	\begin{align}
	\label{CombinerOnlyProblemEquivalent}
	&\underset{\mathbf{W}_\mathrm{RF}, \mathbf{W}_\mathrm{BB}}{\minimize}\hspace{3pt}
	\big|\big| \boldsymbol{\Lambda}_{\bar{\mathrm{y}}}^{1/2} (\mathbf{W}_\mathrm{MMSE}
	- \mathbf{W}_\mathrm{RF} \mathbf{W}_\mathrm{BB}) \big|\big|_\mathcal{F}^2 \nonumber \\
	&\subjectto \mathbf{W}_\mathrm{RF} \in{\mathcal{W}}_\mathrm{RF}.
	\end{align}
	where $\boldsymbol{\Lambda}_{\bar{\mathrm{y}}} = 
	%\frac{\rho}{N_\mathrm{S}}
	\rho\mathbf{H}\mathbf{F}_\mathrm{RF}\mathbf{F}_\mathrm{BB}\mathbf{F}_\mathrm{BB}^\textsf{H}\mathbf{F}_\mathrm{RF}^\textsf{H}\mathbf{H}^\textsf{H} + \sigma_n^2\mathbf{I}_{N_\mathrm{R}}$
	denotes the covariance of the receive array output in (\ref{arrayOutput}).
	In (\ref{CombinerOnlyProblemEquivalent}), the multiplicative term $\boldsymbol{\Lambda}_{\bar{\mathrm{y}}}^{1/2}$ has no element depending on $\mathbf{W}_\mathrm{RF}$ or $\mathbf{W}_\mathrm{BB}$, therefore, it can be removed since it does not affect the solution. 
	Thus, we can solve the combiner design problem in (\ref{CombinerOnlyProblemEquivalent}) as
	\begin{align}
	\label{CombinerDesign}
	&\;\;\;\;\;\;\;\underset{\mathbf{W}_\mathrm{RF}, \mathbf{W}_\mathrm{BB} }{\minimize}\hspace{3pt}
	\big|\big| {\mathbf{W}}_\mathrm{MMSE}
	- \mathbf{W}_\mathrm{RF} {\mathbf{W}}_\mathrm{BB}\big|\big|_\mathcal{F}^2 \nonumber \\
	&\subjectto   \nonumber \\
	&\;\;\;\;\;\;\; \mathbf{W}_\mathrm{RF} \in{\mathcal{W}}_\mathrm{RF} \nonumber \\
	&\;\;\;\;\;\;\; \mathbf{W}_\mathrm{BB} = (\mathbf{W}_\mathrm{RF}^\textsf{H} \boldsymbol{\Lambda}_{\bar{\mathrm{y}}} \mathbf{W}_\mathrm{RF})^{-1}
	(\mathbf{W}_\mathrm{RF}^\textsf{H}\boldsymbol{\Lambda}_{\bar{\mathrm{y}}}\mathbf{W}_\mathrm{MMSE}).
	\end{align}
	{\color{black} The solution to the optimization problem in (\ref{CombinerDesign}) is similar to the precoder design problem in (\ref{PrecoderDesign}) and they can be effectively solved via alternating minimization approach by optimizing each unknown term while fixing the another.} This can be performed by a MATLAB-based algorithm, called Manopt \cite{manopt}.	Note that  (\ref{PrecoderDesign}) and (\ref{CombinerDesign}) do not require a predefined codebook which includes the set of array responses of the receive and transmit arrays. In fact, the optimization problems can be initialized from a random point, i.e., the beamformers with unit-modulus constraint and random phases.

	%%-----------------------------------------------------	
	\begin{figure}[t]
		\centering
		{\includegraphics[draft=false,width=\columnwidth]{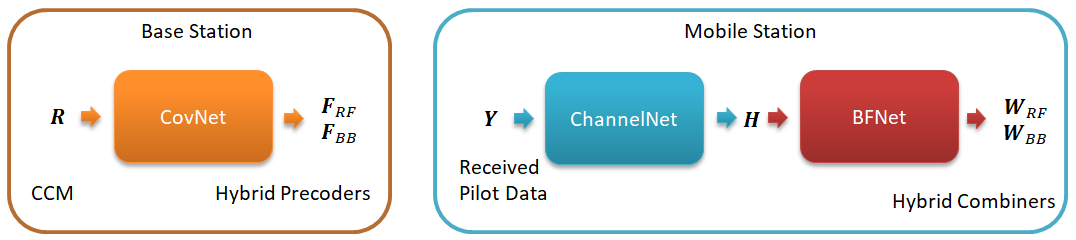} } 
		\caption{The proposed DL framework without instantaneous CSI feedback. The BS feeds the CCM to \textsf{CovNet} to obtain the hybrid precoders $\mathbf{F}_\mathrm{RF}$ and $\mathbf{F}_\mathrm{BB}$. The MS first estimates the channel $\mathbf{H}$ from \textsf{ChannelNet} via the received pilot data $\mathbf{Y}$. Then the hybrid combiners are designed from \textsf{BFNet} by feeding the channel matrix $\mathbf{H}$.  }
		\label{fig_DLFramework}
	\end{figure}
	%%-----------------------------------------------------	
	
	%%-----------------------------------------------------	
	\begin{figure}[t]
		\centering
		{\includegraphics[draft=false,width=\columnwidth,height=.4\textheight]{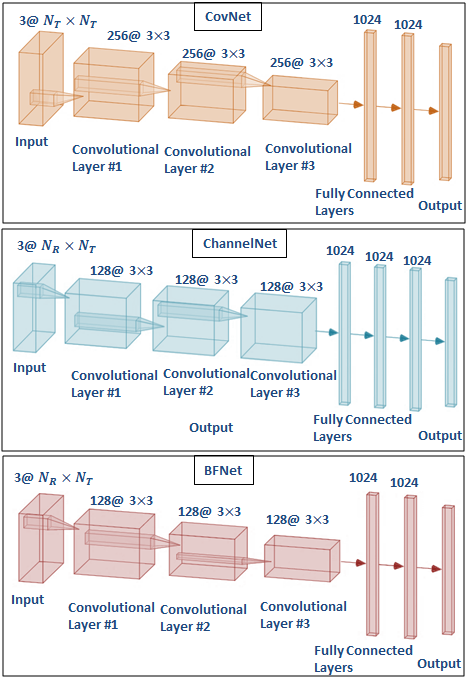} } 
		\caption{The proposed deep network architectures: \textsf{CovNet}, \textsf{ChannelNet} and \textsf{BFNet}.  }
		\label{fig_Networks}
	\end{figure}
	%%-----------------------------------------------------	
	
	\section{Learning-Based Hybrid Beamformer Design}
	\label{sec:Learning}
	In Fig.~\ref{fig_DLFramework}, we demonstrate the proposed DL framework without instantaneous CSI feedback. We introduce three deep network architectures which are shown in Fig.~\ref{fig_Networks}: \textsf{CovNet}, used at the BS only and it learns the channel statistics from $\mathbf{R}$ and obtain the hybrid precoders $\mathbf{F}_\mathrm{RF}$ and $\mathbf{F}_\mathrm{BB}$. \textsf{ChannelNet} and \textsf{BFNet} are placed at the MS only, to estimate the channel $\mathbf{H}$ and construct the hybrid combiners $\mathbf{W}_\mathrm{RF}$ and $\mathbf{W}_\mathrm{BB}$ respectively. In the following, we discuss the details of each deep network architecture.
	
	\subsection{Designing Input Data For The Deep Networks}
	\label{inputData}
	In order to enrich the input features, we feed the networks with three "channel"  with notation $3$@ $D_x \times D_y$ where $D_x$ and $D_y$ denote the 2D sizes of the input. For each "channel", we use real, imaginary, absolute value and the phase of each entry of the input data depending on the application. This approach provides good features for fitting the data in the training state as well as extracting new features inherit in the input~\cite{elbirDL_COMML,elbirQuantizedCNN2019,elbirTVT,deepCNN_ChannelEstimation,elbirIETRSN2019}. In particular, we denote the input for \textsf{CovNet} as $\mathbf{X}_\mathbf{R} \in \mathbb{R}^{N_\mathrm{T}\times N_\mathrm{T}\times 3}$ whose $(i,j)$-th entry of the first, second and the third "channel" is given by $[[\mathbf{X}_\mathbf{R}]_{:,:,1}]_{i,j} = \operatorname{Re}\{ [\mathbf{R}]_{i,j}\}$, $[[\mathbf{X}_\mathbf{R}]_{:,:,2}]_{i,j} = \operatorname{Im} \{[\mathbf{R}]_{i,j}\}$ and $[[\mathbf{X}_\mathbf{R}]_{:,:,3}]_{i,j} =  \angle\{[\mathbf{R}]_{i,j}\}$, respectively.	For \textsf{ChannelNet}, the input is denoted as $\mathbf{X}_\mathbf{Y}\in \mathbb{R}^{N_\mathrm{R}\times N_\mathrm{T}\times 3}$ and, similarly we have $[[\mathbf{X}_\mathbf{Y}]_{:,:,1}]_{i,j} = \operatorname{Re}\{ [\mathbf{Y}]_{i,j}\}$, $[[\mathbf{X}_\mathbf{Y}]_{:,:,2}]_{i,j} = \operatorname{Im} \{[\mathbf{Y}]_{i,j}\}$ and $[[\mathbf{X}_\mathbf{Y}]_{:,:,3}]_{i,j} =  |[\mathbf{Y}]_{i,j}|$. Finally, the input for \textsf{BFNet} is given by $\mathbf{X}_\mathbf{H}\in \mathbb{R}^{N_\mathrm{R}\times N_\mathrm{T}\times 3}$ where  $[[\mathbf{X}_\mathbf{H}]_{:,:,1}]_{i,j} = \operatorname{Re}\{ [\mathbf{H}]_{i,j}\}$, $[[\mathbf{X}_\mathbf{H}]_{:,:,2}]_{i,j} = \operatorname{Im}\{ [\mathbf{H}]_{i,j}\}$ and $[[\mathbf{X}_\mathbf{H}]_{:,:,3}]_{i,j} =  |[\mathbf{H}]_{i,j}|$ respectively. We observe, through simulations, that the angular values provide better features and training performance for covariance data whereas the absolute value is more appropriate for $\mathbf{Y}$ and $\mathbf{H}$~\cite{elbirIETRSN2019}. Hence, the third "channel" for $\mathbf{X}_\mathbf{R}$ is selected as the angle information whereas the absolute value is used for the third "channel" of $\{\mathbf{X}_\mathbf{Y},\mathbf{X}_\mathbf{H}\}$.

	\subsection{Labeling The Deep Networks}
	\label{labelData}
	We start by constructing the labels of \textsf{CovNet} which is the hybrid precoders $\mathbf{F}_\mathrm{RF}$ and $\mathbf{F}_\mathrm{BB}$. Hence, we represent the output label of \textsf{CovNet} by $\mathbf{z}_\mathbf{R}$ as
	\begin{align}
	\label{zR}
	\mathbf{z}_\mathbf{R} = [\mathrm{vec}\{\angle \mathbf{F}_\mathrm{RF}  \}^\textsf{T},
	\mathrm{vec}\{\operatorname{Re}\{ \mathbf{F}_\mathrm{BB}\} \}^\textsf{T}, \mathrm{vec}\{\operatorname{Im}\{ \mathbf{F}_\mathrm{BB}\}\}^\textsf{T} ]^\textsf{T},
	\end{align}
	which is an $N_\mathrm{RF}\big(N_\mathrm{T}  + 2N_\mathrm{S} \big) \times 1$ real-valued vector. For \textsf{ChannelNet}, we represent the labels by $\mathbf{z}_\mathbf{Y}$ as 
	\begin{align}
	\label{zY}
	\mathbf{z}_{\mathbf{Y}} = [\mathrm{vec}\{\operatorname{Re}\{\mathbf{H}\}\}^\textsf{T} , \mathrm{vec}\{\operatorname{Im}\{\mathbf{H}\}\}^\textsf{T} ]^\textsf{T},
	\end{align}
	which is a real-valued vector of size $2N_\mathrm{R}N_\mathrm{T}\times 1$. Finally, the output label of \textsf{BFNet} is, similar to \textsf{CovNet}, given by $\mathbf{z}_\mathbf{H} \in \mathbb{R}^{N_\mathrm{RF}\big(N_\mathrm{R}  + 2N_\mathrm{S} \big) }$ as
	\begin{align}
	\label{zH}
	\mathbf{z}_\mathbf{H} = [\mathrm{vec}\{\angle \mathbf{W}_\mathrm{RF}  \}^\textsf{T},
	\mathrm{vec}\{\operatorname{Re}\{ \mathbf{W}_\mathrm{BB}\} \}^\textsf{T}, \mathrm{vec}\{\operatorname{Im}\{ \mathbf{W}_\mathrm{BB}\} ]^\textsf{T}.
	\end{align}

	\subsection{Network Architectures and Training}
	\label{Training}
	The deep networks in Fig.~\ref{fig_Networks}, \textsf{CovNet, ChannelNet} and \textsf{BFNet} have the input-output pairs as $\{\mathbf{X}_\mathbf{R},\mathbf{z}_\mathbf{R} \}$, $\{\mathbf{X}_\mathbf{Y},\mathbf{z}_\mathbf{Y} \}$
	and $\{\mathbf{X}_\mathbf{H},\mathbf{z}_\mathbf{H} \}$	respectively. For each network, we use three convolutional layers with kernel size of $3\times 3$. While \textsf{CovNet} has 256 convolutional filters, \textsf{ChannelNet} and \textsf{BFNet} have 128 filters. In addition, \textsf{CovNet} and \textsf{BFNet} have two pooling layers, which reduce the dimension by two, after the first two convolutional layers whereas \textsf{ChannelNet} has no pooling layer. There are two fully connected layers in \textsf{CovNet} and \textsf{BFNet} and three fully connected layers are placed in \textsf{ChannelNet}. There are dropout layers with a $50\%$ probability after each fully connected layer in each network. The output layer of all networks are the regression layer with proper size depending on the application as discussed in Section~\ref{labelData}. While the other network architectures with different parameters are also possible, the presented network parameters are one possible solution to obtain good performance for the considered problem. We have obtained the network parameters from a hyperparameter tuning process providing the best performance for the considered scenario \cite{elbirDL_COMML,elbirQuantizedCNN2019,elbirIETRSN2019,elbirTVT,deepCNN_ChannelEstimation}.

	The proposed deep networks are realized and trained in MATLAB on a PC with a single GPU and a  768-core processor. We have summarized the algorithmic steps for training data generation in Algorithm~\ref{alg:algorithmTraining}. We have used the stochastic gradient descent algorithm with momentum 0.9  and  updated the network parameters with learning rate $0.0005$ and mini-batch size of $128$ samples. Then, we have reduced the learning rate by the factor of $0.9$ after each 20 epochs. We also applied a stopping criteria during training such that the training terminates if the validation accuracy does not improve in three consecutive epochs.  To train the proposed CNN structures, $N=100$ different scenarios are realized for $G=200$ as in Algorithm~\ref{alg:algorithmTraining}. For each scenario, we generated a channel matrix (together with the corresponding covariance matrix) and received pilot signal where synthetic additive noise is added to the training data on the CCM, channel matrix and the received pilot signal which are defined by SNR$_{\mathbf{R}}$, SNR$_{\mathbf{H}}$ and SNR$_{\bar{\mathbf{N}}}$ respectively\footnote{
		In the simulations, we have used four SNR definitions, all of which are characterized by AWGN. 1) $\{$SNR$_{\bar{\mathbf{N}}}$, SNR$_{\bar{\mathbf{N}}-\mathrm{TEST}}\}$: SNR on the  signal in (\ref{receivedSignalPilotMod}) when the pilot signals are received in the preamble for training and test stage respectively. 2)  SNR$_{\mathbf{H}}$: SNR on the channel matrix to obtain the corrupted channel data in training. 3)  SNR$_{\mathbf{R}}$ : SNR on the channel covariance matrix to obtain the corrupted channel covariance data in training. 4) Finally, we use	the term "SNR" on the received signal in (\ref{arrayOutput}) (not in the preamble) for hybrid beamforming process.}. In the training process we use multiple SNR$_{\mathbf{R}}$, SNR$_{\mathbf{H}}$ and SNR$_{\bar{\mathbf{N}}}$ values to make the networks robust against corrupted input characteristics~\cite{elbirDL_COMML,elbirQuantizedCNN2019}. Hence we use  SNR$_{\bar{\mathbf{N}}} = \{20, 30, 40\}$ dB, SNR$_{\mathbf{H}} =\{15,20,25\}$ dB  and SNR$_{\mathbf{R}} =\{20, 25, 30\}$ dB. Hence, we define SNR$_{\mathbf{R}} = 20\log_{10}(\frac{|[\mathbf{R}]_{i,j}|^2}{\sigma_{\mathbf{R}}^2})$, SNR$_{\mathbf{H}} = 20\log_{10}(\frac{|[\mathbf{H}]_{i,j}|^2}{\sigma_{\mathbf{H}}^2})$ and SNR$_{\bar{\mathbf{N}}} = 20\log_{10}(\frac{|[ \mathbf{H} \bar{\mathbf{F}}\bar{\mathbf{S}}]_{i,j}|^2}{\sigma_{\bar{\mathbf{N}}}^2})$ where $\sigma_{\bar{\mathbf{N}}}^2,\sigma_{\mathbf{H}}^2,\sigma_{\mathbf{R}}^2$ are the variance of AWGN corresponding to the input data. As a result,  180000 input-output pairs are generated for training. In the training process, $80\%$ and $20\%$ of all generated data  are selected as the training and validation datasets, respectively.  For the prediction process, we have generated a test data which is separately generated by adding noise on received pilot signal with  SNR$_{\bar{\mathbf{N}}-\mathrm{TEST}}$. Note that this allows us to further corrupt the input data and test the network against deviations in the input data which can resemble the changes in the mm-Wave channel \cite{coherenceTimeRef}. 
	%	In particular, we define the corresponding SNR as  SNR$_{\mathbf{H}-\mathrm{TEST}} = 20\log_{10}(\frac{|[\mathbf{H}]_{i,j}|^2}{\sigma_{\mathbf{H}-\mathrm{TEST}}^2})$. 

	%-------------------------------------------------------------------------------------------------
	\begin{algorithm}[t]
		\begin{algorithmic}[1]
			\caption{Training data generation for \textsf{CovNet, ChannelNet} and \textsf{BFNet}. }
			\Statex {\textbf{Input:} $N$,  $G$,  SNR$_{{\mathbf{H}}}$, SNR$_{\bar{\mathbf{N}}}$  }.
			\Statex{\textbf{Output:} Training datasets for the networks in Fig.~\ref{fig_Networks}: $\mathcal{D}_{\textsf{CovNet}}$, $\mathcal{D}_{\textsf{ChannelNet}}$ and $\mathcal{D}_{\textsf{BFNet}}$.}
			\label{alg:algorithmTraining}
			%			\Statex
			\State Generate channel covariance matrix realizations $\{\mathbf{R}^{(n)}\}_{n=1}^N$.
			\State Generate channel realizations $\{\mathbf{H}^{(n)}\}_{n=1}^N$.
			%			\State Generate $\{\bar{\mathbf{S}}^{(q)}[m]\}_{q = 1}^Q$ for $m \in \mathcal{M}$.
			\State Initialize with $t=1$ while the dataset length is $T=NG$.
			%		\State \mathbf{while}  $t\leq T$ \mathbf{do}
			\State   \textbf{for}  $1 \leq n \leq N$ \textbf{do}
			\State  \indent \textbf{for}  $1 \leq g \leq G$ \textbf{do}
			\State \indent  $[\tilde{\mathbf{H}}^{(n,g)}]_{i,j} \sim \mathcal{CN}([\mathbf{H}^{(n)}]_{i,j},\sigma_{\mathbf{H}}^2)$.
			\State \indent  $[\tilde{\mathbf{R}}^{(n,g)}]_{i,j} \sim \mathcal{CN}([\mathbf{R}^{(n)}]_{i,j},\sigma_{\mathbf{R}}^2)$.
			\State \indent The BS transmits the pilot and it is received as
			\begin{align}
			\bar{\mathbf{Y}}^{(n,g)} = \bar{\mathbf{W}}^{\textsf{H}} \mathbf{H}^{(n,g)}\bar{\mathbf{F}} + \tilde{\mathbf{N}}^{(n,g)}. \nonumber 
			\end{align}
			%			\indent where $\mathbf{N}[m]$ is generated with SNR$_{\bar{\mathbf{N}}}$.
			\State \indent Construct ${\mathbf{Y}}^{(n,g)}$ from (\ref{Y}) by using $\bar{\mathbf{Y}}^{(n,g)}$.
			\State \indent  The BS designs the precoders ${\mathbf{F}}_{\mathrm{RF}}^{(n,g)}$ and ${\mathbf{F}}_{\mathrm{BB}}^{(n,g)}$ by \par \indent using  $\mathbf{R}^{(n,g)}$ from (\ref{PrecoderDesign}).
			\State \indent  The MS finds  ${\mathbf{W}}_{\mathrm{RF}}^{(n,g)}$ and ${\mathbf{W}}_{\mathrm{BB}}^{(n,g)}$   by solving  (\ref{CombinerDesign}).
			\State \indent Construct input $\mathbf{X}_\mathbf{R}^{(t)}$, $\mathbf{X}_\mathbf{Y}^{(t)}$ and $\mathbf{X}_\mathbf{H}^{(t)}$ from $\tilde{\mathbf{R}}^{(n,g)}$, \par \indent  ${\mathbf{Y}}^{(n,g)}$ and  $\tilde{\mathbf{H}}^{(n,g)}$, respectively, as in Section~\ref{inputData}.
			\State \indent Construct output $\mathbf{z}_\mathbf{R}^{(t)}$, $\mathbf{z}_\mathbf{Y}^{(t)}$ and $\mathbf{z}_\mathbf{H}^{(t)}$ from \par \indent $\{{\mathbf{F}}_{\mathrm{RF}}^{(n,g)},{\mathbf{F}}_{\mathrm{BB}}^{(n,g)}\}$,  $\mathbf{H}^{(n,g)}$ and $\{{\mathbf{F}}_{\mathrm{RF}}^{(n,g)},{\mathbf{F}}_{\mathrm{BB}}^{(n,g)}\}$, \par \indent respectively, from  in (\ref{zR}), (\ref{zY}) and (\ref{zH}).

			\State \indent $t = t+1$.
			
			\State \indent\textbf{end for} $g$,	
			\State \textbf{end for} $n$,
			\State $\mathcal{D}_{\textsf{CovNet}} = \big((\mathbf{X}_{\mathbf{R}}^{(1)}, \mathbf{z}_\mathbf{R}^{(1)}),\dots, (\mathbf{X}_{\mathbf{R}}^{(T)}, \mathbf{z}_\mathbf{R}^{(T)})\big).$
			\State $\mathcal{D}_{\textsf{ChannelNet}} = \big((\mathbf{X}_\mathbf{Y}^{(1)}, \mathbf{z}_\mathbf{Y}^{(1)} ),\dots, (\mathbf{X}_\mathbf{Y}^{(T)}, \mathbf{z}_\mathbf{Y}^{(T)} )\big).$
			\State $\mathcal{D}_{\textsf{BFNet}} = \big((\mathbf{X}_\mathbf{H}^{(1)}, \mathbf{z}_\mathbf{H}^{(1)}),\dots, (\mathbf{X}_\mathbf{H}^{(T)}, \mathbf{z}_\mathbf{H}^{(T)})\big).$
		\end{algorithmic}
	\end{algorithm}
	%------------------------------------------------------------------------------------------------

	{\color{black}
		
		\subsection{Online Deployment of the Proposed DL Approach}
		\label{sec:OnlineDeploy}
		The adaptation of the DL network to the changes in the propagation environment is very important since the offline training cannot include all possible channel characteristics. In this part, we investigate the online performance of the proposed DL approach for channel estimation. In particular, \textsf{ChannelNet} is trained offline as described in Algorithm~\ref{alg:algorithmTraining}. Then, it is deployed online where the instantaneous channel parameters such as $\alpha_{l,r}$, $\theta_{l,r}$ and $\phi_{l,r}$ change due to the motion of MS. Let $\mathbf{Y}_{t}$ be the received pilot signal at time $t$, which is constructed as in the $\{8,9\}$-th steps of Algorithm~\ref{alg:algorithmTraining}. Since in the online stage the channel matrix is unknown, $\mathbf{Y}_t$ is used to estimate the channel by both \textsf{ChannelNet} and also an analytical channel estimation technique such as angle domain channel estimation (ADCE) approach proposed in~\cite{angularDomainCE_feifei}, which has close-to-CRB (Cramer-Rao lower bound) performance. Then, the DL network is updated if its performance is poor, otherwise it continues with its current form. Let us define the error metric at time $t$ as $\eta^{(t)}$ which is defined as 
		\begin{align}
		\eta^{(t)} = || \hat{\mathbf{H}}_t^\mathrm{DL} - \hat{\mathbf{H}}^\mathrm{temp}   ||_\mathcal{F}/(N_\mathrm{T} N_\mathrm{R}), 
		\end{align}
		where $\hat{\mathbf{H}}^\mathrm{temp} $ is the last estimated channel matrix via ADCE. Note that the channel estimation via ADCE is only performed at the beginning of the online deployment and when the network is updated. By doing so, the channel estimation complexity is reduced thanks to the low computation time of DL network. The network updates its parameters if
		\begin{align}
		\eta^{(t)} \geq \zeta,
		\end{align}
		holds for some threshold parameter $\zeta$ which determines how frequently the network is updated.  If $\eta^{(t)} < \zeta$, then it means that \textsf{ChannelNet} performs satisfactorily and it continues to work without update. Let us denote network parameters at time $t$ as $\boldsymbol{\Pi}_t$, then, we update only the higher layers (i.e., the fully connected layers) of $\boldsymbol{\Pi}_t$, which are more environment-dependent. The lower layers (i.e., convolutional layers) are kept intact or \textit{frozen} because they generally behave like problem-dependent~\cite{pan2010domain}. In Algorithm~\ref{alg:OnlineDeploy}, we describe the steps for online deployment. Here SNR$_{\mathbf{Y}} = 20\log_{10}(\frac{|[\mathbf{Y}]_{i,j}|^2}{\sigma_{\mathbf{Y}}^2})$ is used to introduce noisy inputs for $\boldsymbol{\Pi}_t$ to provide robustness.
		
		During online deployment, the complexity of the training is low due to the small size of dataset. Furthermore, the online training is only performed if the propagation environment changes significantly. Hence, the proposed DL approach does not need to be re-trained for each time instance, which lowers the computational complexity.

		%-------------------------------------------------------------------------------------------------
		\begin{algorithm}[t]
			{\color{black}
				\begin{algorithmic}[1]
					\caption{Online deployment for \textsf{ChannelNet}. }
					\Statex {\textbf{Input:} $\mathbf{Y}_t$, $\boldsymbol{\Pi}_t$, SNR$_{{\mathbf{Y}}}$, $\zeta$, $G_\mathrm{Y}$}.
					\Statex	{\textbf{Output:} $\hat{\mathbf{H}}_t^\mathrm{DL}$.}
					\label{alg:OnlineDeploy}
					\State Start with $t=0$. 
					\State Estimate $\mathbf{H}_t$ via  $\boldsymbol{\Pi}_t$ and ADCE as $\hat{\mathbf{H}}_t^{\mathrm{DL}}$ and  $\hat{\mathbf{H}}_t^{\mathrm{AD}}$.
					\State Set $\hat{\mathbf{H}}^\mathrm{temp} = \hat{\mathbf{H}}_t^{\mathrm{AD}}$.
					\State Compute the error metric $\eta^{(t)}$.
					%			 $\boldsymbol{\Pi}_t$ and ADCE as $\eta_\mathrm{DL}^{(t)}$ and $\eta_\mathrm{AD}^{(t)}$.
					\State \textbf{while} $\eta^{(t)}  < \zeta$, \textbf{do}
					\State \indent  Estimate $\mathbf{H}_t$ via $\boldsymbol{\Pi}_t$  as $\hat{\mathbf{H}}_t^{\mathrm{DL}}$.
					\State \indent $\boldsymbol{\Pi}_{t+1}= \boldsymbol{\Pi}_{t}$.
					\State \indent Compute $\eta^{(t)}$.
					\State \indent $t \leftarrow t + 1$.
					\State \textbf{else}
					\State \indent Estimate $\mathbf{H}_t$ via  ADCE as  $\hat{\mathbf{H}}_t^{\mathrm{AD}}$.
					\State \indent Set $\hat{\mathbf{H}}^\mathrm{temp} = \hat{\mathbf{H}}_t^{\mathrm{AD}}$.
					\State \indent \textbf{for}  $1 \leq g \leq G_\mathrm{Y}$ \textbf{do}
					\State \indent \indent $[\mathbf{Y}_t^{(g)}]_{i,j} \sim \mathcal{CN}([\mathbf{Y}_t]_{i,j}, \sigma_\mathbf{Y}^2)$,
					\State \indent \indent $\mathbf{z}_t^{(g)} = [\mathrm{vec}\{\operatorname{Re}\{\hat{\mathbf{H}}_t^{\mathrm{AD}}\}\}^\textsf{T}, \mathrm{vec}\{\operatorname{Im}\{\hat{\mathbf{H}}_t^{\mathrm{AD}}\}\}^\textsf{T}]$,
					\State \indent \textbf{end}
					\State \indent Construct the online update dataset $\mathcal{D}_{t-\mathrm{online}}$ where \par \indent  $\mathcal{D}_{t-\mathrm{online}}^{(g)} = \big(\mathbf{Y}_t^{(g)}, \mathbf{z}_t^{(g)} \big)$.
					\State \indent Freeze the convolutional layers of $\boldsymbol{\Pi}_t$.
					\State \indent Update $\boldsymbol{\Pi}_t$ with $\mathcal{D}_\mathrm{online}$ and obtain $\boldsymbol{\Pi}_{t+1}$.
					\State \indent Estimate  $\mathbf{H}_t$ via $\boldsymbol{\Pi}_{t+1}$ and refine estimate $\hat{\mathbf{H}}_t^{\mathrm{DL}}$.
					
					\State \indent $t \leftarrow t + 1$.
					
					\State \textbf{end}
				\end{algorithmic}
			}
		\end{algorithm}
		%------------------------------------------------------------------------------------------------
		
		\subsection{Computational Complexity and Power Consumption}
		\label{sec:CComp}
		The computational complexity of the proposed DL approach has two main parts, namely, online prediction and offline training. While the online complexity has simple expressions, the complexity analysis of the offline training is still an open issue due to the complex implementation of backpropagation process involved during training. Therefore, we only consider the complexity of online prediction stage. 
		
		For a deep neural network with $L_\mathrm{C}$ convolutional layers and $L_\mathrm{F}$ fully connected layers~\cite{vggRef}, the total time complexity of convolutional layers is $\mathcal{O}( \sum_{l=1}^{L_\mathrm{C}}D_x^{(l)}D_y^{(l)} b_x^{(l)}b_y^{(l)}c_\mathrm{CL}^{(l-1)}c_\mathrm{CL}^{(l)}   )$ where $D_x^{(l)}, D_y^{(l)}$ are the convolutional kernel size, $ b_x^{(l)},b_y^{(l)}$ are the 2D output size of the $l$-th convolutional layer and $c_\mathrm{CL}^{(l)}$ is the number of filters of the $l$-th layer. The total complexity of the fully connected layers is $\mathcal{O} (\sum_{l=1}^{L_\mathrm{F}}b_x^{(l)} b_y^{(l)} c_\mathrm{F}^{(l)}  ) $ where $c_\mathrm{FL}^{(l)}$ is the number of units in the fully connected layer. The comparison of computational complexity for a DL  network with a conventional analytical method is not fair due to the use of different processing units, i.e., DL benefits the use of GPU, while not all analytical method are implementable via GPU~\cite{vggRef}. For this reason, we compare the computation times of both algorithms in the simulations section.
		
		Next, we compare the power consumption of the conventional systems and the DL-based methods. For a conventional mm-Wave system with transmit power $P_\mathrm{T}=1$ W, total power consumption can be given as
		\begin{align}
		P_\mathrm{tot} = P_\mathrm{T} + N_\mathrm{RF}P_\mathrm{RF} + N_\mathrm{T}N_\mathrm{RF} P_\mathrm{PS} + P_\mathrm{BB},
		\end{align}
		where $ P_\mathrm{RF} $, $P_\mathrm{PS}$ and $P_\mathrm{BB}$ are the power consumption of an RF chain, phase shifter and baseband processing respectively. Furthermore, they are given approximately as  $ P_\mathrm{RF} =250$ mW, $P_\mathrm{PS}=40$ mW and $P_\mathrm{BB}=200$ mW, for which $P_\mathrm{tot}$ becomes approximately $P_\mathrm{tot} = 22.68$ W for $N_\mathrm{T}=128$ and $N_\mathrm{RF}=4$~\cite{mimoChannelModel2}. While there is no commercial DL-based hardware for mm-Wave system configuration, there exist some processing units that can run DL methods effectively such as Intel Movidius~\cite{powerConsumptionDL2} which has power consumption of approximately $500$ mW~\cite{powerConsumptionDL,powerConsumptionDL2}. This shows that DL-based methods can be a promising candidate for next generation wireless communication systems providing reasonable performance and lower computation complexity.	}
	
	%	{\color{black} Next, we examine the power consumption of the proposed DL approach in comparison with the conventional systems. For a conventional mm-Wave system with transmit power $P_\mathrm{T}=1$ W, total power consumption can be given as
	%		\begin{align}
	%		P_\mathrm{tot} = P_\mathrm{T} + N_\mathrm{RF}P_\mathrm{RF} + N_\mathrm{T}N_\mathrm{RF} P_\mathrm{PS} + P_\mathrm{BB},
	%		\end{align}
	%		where $ P_\mathrm{RF} $, $P_\mathrm{PS}$ and $P_\mathrm{BB}$ are the power consumption of an RF chain, phase shifter and baseband processing respectively. These quantities can be given approximately as  $ P_\mathrm{RF} =250$ mW, $P_\mathrm{PS}=40$ mW and $P_\mathrm{BB}=200$ mW, for which $P_\mathrm{tot}$ becomes approximately $P_\mathrm{tot} = 22.68$ W for $N_\mathrm{T}=128$ transmit antennas and $N_\mathrm{RF}=4$ RF chains~\cite{mimoChannelModel2}. While there is no commercial DL-based hardware for mm-Wave system configuration, there exist some processing units that can run DL methods effectively such as Intel Movidius~\cite{powerConsumptionDL2} which has power consumption of approximately $500$ mW~\cite{powerConsumptionDL,powerConsumptionDL2}. This shows that DL-based methods can be a promising candidate for next generation wireless communication systems providing reasonable performance and lower computation complexity.}

	\section{Numerical Simulations}
	\label{sec:Sim}
	In this section, we have evaluated the performance of the proposed approach (called SDHB: statistical deep hybrid beamforming) through several experiments. We compare the proposed approach with both statistical and non-statistical hybrid beamforming techniques such as SHB \cite{widebandHBWithoutInsFeedback} {\color{black}(which solves (\ref{Frf_suboptimum}))}, PE-HB \cite{sohrabiNarrowband} and deep learning-based hybrid beamforming (DLHB) \cite{elbirDL_COMML} as well as the MO algorithm~\cite{hybridBFAltMin}. Note that the performance of the MO algorithm constitutes an upper bound for DLHB/SDHB since the network labels of DLHB/SDHB are obtained by MO. Therefore, DLHB/SDHB cannot perform better than MO. We further evaluate the performance of the fully digital beamforming performance as a benchmark. In order to compare the channel estimation performance of \textsf{ChannelNet}, we implement the SF-CNN algorithm~\cite{deepCNN_ChannelEstimation} with the same network parameters and feed with the same input, i.e., the initial channel estimates.
	
	Throughout the simulations, we consider a single-user massive MIMO system with $N_\mathrm{RF}=4$ RF chains for $N_\mathrm{T}=128$ and $N_\mathrm{R}=16$ antennas. The antennas are deployed with half wavelength spacing at $f_c = 60$ GHz. We assume, unless stated otherwise, there are  $L=5$ clusters of all transmit and receive paths which are uniform randomly selected from the interval $\{\phi,\theta\} \in [-\pi,\pi]$ with angular spread of $5^\circ$. In preamble stage, the transmitter emits only one beam by using a single RF chain while, at the receiver, all of the RF chains are active. The transmit and received beams are formed by selecting  $\bar{\mathbf{F}}$ and $\bar{\mathbf{W}}$ as $N_\mathrm{T}\times N_\mathrm{T}$ and $N_\mathrm{R}\times N_\mathrm{R}$ DFT matrices respectively \cite{deepCNN_ChannelEstimation}.

	%%-----------------------------------------------------
	\begin{figure}[t]
		\centering
		{\includegraphics[draft=false,width=\columnwidth,height=.7\columnwidth]{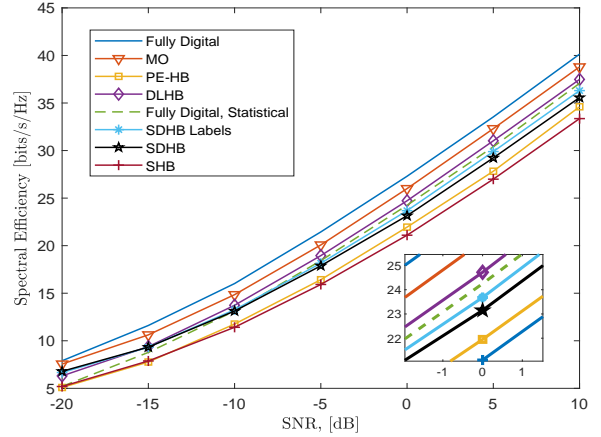} } 
		\caption{Spectral efficiency versus SNR. $N_\mathrm{T} = 128$, $N_\mathrm{R}=16$ and  SNR$_{\bar{\mathbf{N}}-\mathrm{TEST}}=10$ dB.   }
		\label{fig_SNR_Rate}
	\end{figure}
	%	%%-----------------------------------------------------

	In Fig.~\ref{fig_SNR_Rate}, we present the spectral efficiency with respect to SNR when SNR$_{\bar{\mathbf{N}}-\mathrm{TEST}}=10$ dB. As expected, we see that the non-statistical approaches (MO, PE-HB, DLHB) perform superior than the statistical approaches (SHB, SDHB) including the fully-digital beamforming performance, thanks to the available instantaneous CSI knowledge. In statistical approaches, the CCM does not fully reflect the instantaneous channel data which changes in time due to the parameters such as the channel gain $\alpha_{l,r}$. We can also see that DL-based approaches outperform the non-DL techniques such as PE-HB and SHB. In particular, the proposed approach SDHB closely follows the fully-digital beamforming for statistical case. The outperformance of SDHB  can be attributed to the use of optimum hybrid beamformers as labels, which are obtained by the MO algorithm~\cite{hybridBFAltMin} whereas SHB simply takes the phases of $\mathbf{F}^\mathrm{opt}$ which is sub-optimum. Furthermore, as will be demonstrated later, the time complexity of the proposed DL-based approach is much lower as compared to the MO, thus making it a very efficient algorithm. 
	
	%	In comparison with SHB, SDHB provides better performance due to the manifold optimization process taken place when obtaining the analog beamformers. In SHB, the analog hybrid beamformer is obtained by simply taking the phases of the unconstrained beamformer which is a sub-optimum process. 

\begin{figure}[t]
		\centering
		{\includegraphics[draft=false,width=\columnwidth,height=.7\columnwidth]{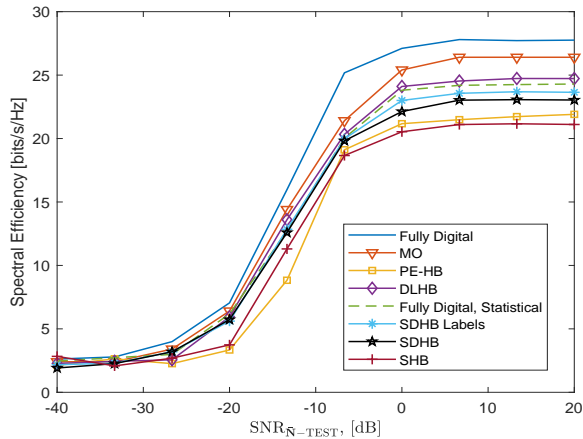} } 
		\caption{Spectral efficiency versus SNR$_{\bar{\mathbf{N}}-\mathrm{TEST}}$ when SNR$=0$ dB.   }
		\label{fig_SNRonPreamble_Rate}
	\end{figure}

	In Fig.~\ref{fig_SNRonPreamble_Rate}, the robustness of the algorithms is investigated with respect to the estimated channel data. Hence, for all of the beamforming algorithms, we use the channel matrix estimated by \textsf{ChannelNet} when the received pilot data is corrupted by noise determined by SNR$_{\bar{\mathbf{N}}-\mathrm{TEST}}$. Note that the noise introduced by SNR$_{\bar{\mathbf{N}}-\mathrm{TEST}}$ only affects the combiner design (not the precoder design) performance of the statistical approaches (SHB and SDHB) since they only use estimated channel data in the combiner design stage. We can see from Fig.~\ref{fig_SNRonPreamble_Rate} that all of the algorithms reach their maximum performance after SNR$_{\bar{\mathbf{N}}-\mathrm{TEST}}\geq 0$ dB. In particular, SDHB has more robust performance than SHB and performs very close to the fully-digital beamformer. This observation states that the algorithms require at least approximately SNR$_{\bar{\mathbf{N}}-\mathrm{TEST}}=-5$ dB noise level for  sufficient channel estimate in this setting.
	
\begin{figure}[t]
		\centering
		{\includegraphics[draft=false,width=\columnwidth,height=.7\columnwidth]{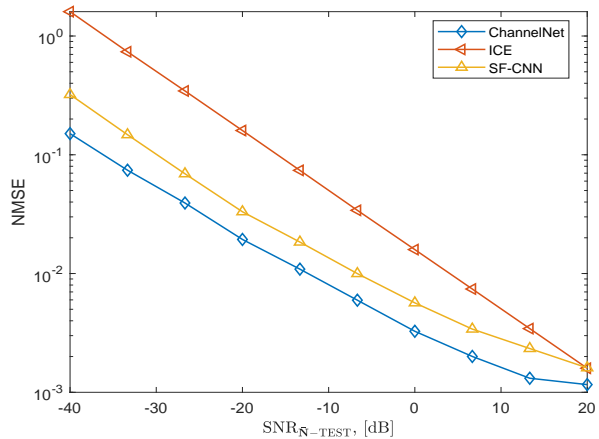} } 
		\caption{Normalized MSE versus SNR$_{\bar{\mathbf{N}}-\mathrm{TEST}}$ when SNR$=0$ dB.   }
		\label{fig_SNRonPreamble_NMSE}
	\end{figure}

	For the same experiment, we also examine the channel estimation performance in terms of normalized MSE in Fig.~\ref{fig_SNRonPreamble_NMSE}. We see that both \textsf{ChannelNet} and SF-CNN  outperform the  initial channel estimate obtained from the received pilots. We also see that  \textsf{ChannelNet} outperforms SF-CNN which cannot do well, especially for high SNR$_{\bar{\mathbf{N}}-\mathrm{TEST}}$. The poor performance of SF-CNN is because SF-CNN uses several convolutional layers and no fully connected layer. While convolutional layers are good at extracting new features from the input, fully connected layers are more powerful in terms of mapping the input data to the output~\cite{vggRef}.
	%%-----------------------------------------------------
	%	%%-----------------------------------------------------
	Thanks to the three fully connected layers in \textsf{ChannelNet}, it achieves much less NMSE than SF-CNN. Another disadvantage of SF-CNN~\cite{deepCNN_ChannelEstimation} is that SF-CNN works properly in the prediction stage only if the SNR in the prediction stage is the same as of the training stage, i.e., SNR$_{\bar{\mathbf{N}}}$ = SNR$_{\bar{\mathbf{N}}-\mathrm{TEST}}$ is required.  Such a requirement is not needed in our DL framework thanks to the use of multiple SNR$_{\bar{\mathbf{N}}}$ levels during training. We see that the performance of both \textsf{ChannelNet} and SF-CNN makes out as SNR$_{\bar{\mathbf{N}}-\mathrm{TEST}}$ increases, especially for SNR$_{\bar{\mathbf{N}}-\mathrm{TEST}}>10$ dB. This is due to the lack of precision of the deep networks, which are biased estimators in nature. Higher precision can still be obtained if larger number of units in the network layers are used with less training data size. However, this will cause the network memorize the input data so that the network will not function if input data differs from the ones used  in the training data set~\cite{elbirIETRSN2019}. This fact suggests that the training data should not include too much indistinguishable (noisy) data to provide good precision.  
	%The current network architecture is obtained through an optimization process so that it performs the best for the given scenario while the other architectures are still possible, and it is not the main focus of this work. Nevertheless, we have shown that the proposed \textsf{ChannelNet} architecture provides very good channel estimation performance even when the test data  is completely new and different than the training data. 
	
	%%-----------------------------------------------------
	\begin{figure}[t]
		\centering
		{\includegraphics[draft=false,width=\columnwidth,height=.7\columnwidth]{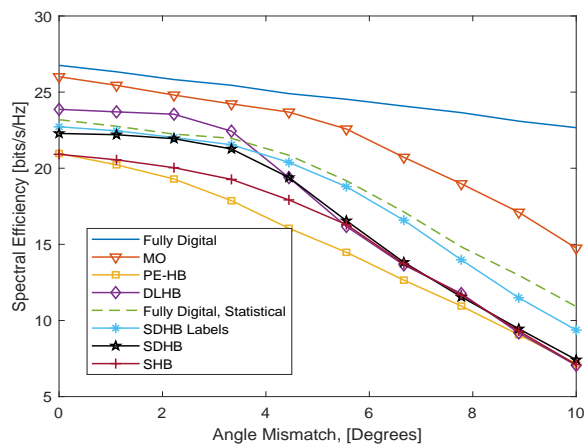} } 
		\caption{Spectral efficiency versus angle mismatch when SNR$=0$ dB and SNR$_{\bar{\mathbf{N}}-\mathrm{TEST}}=10$ dB.   }
		\label{fig_SNR_AngleMismatch_Rate}
	\end{figure}
	\begin{figure}[t]
		\centering
		{\includegraphics[draft=false,width=\columnwidth,height=.7\columnwidth]{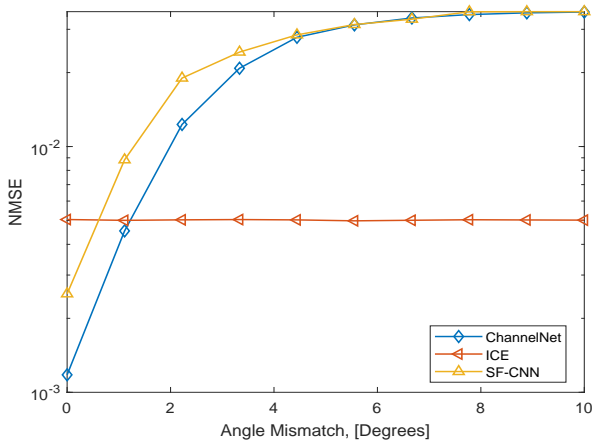} } 
		\caption{Channel estimation NMSE versus angle mismatch when SNR$=0$ dB and SNR$_{\bar{\mathbf{N}}-\mathrm{TEST}}=10$ dB.   }
		\label{fig_SNR_AngleMismatch_NMSE} 
	\end{figure}
	
	%	%%-----------------------------------------------------

	In Fig~\ref{fig_SNR_AngleMismatch_Rate}, we present the performance of the algorithms when there is an angular mismatch in the received path angles between the channel matrix used in the training data and the test data. Specifically, the trained networks are fed with the channel data that is generated by introducing angular mismatch in the AOA/AOD angles of all received paths of the channel matrices used in the training data. In a similar way, we also introduce angular mismatch in the CCM data that is used for \textsf{CovNet} and SHB. We observe from Fig.~\ref{fig_SNR_AngleMismatch_Rate} that as the standard deviation of the angular mismatch increases, expectedly,  the spectral efficiency performance becomes poorer due to the loss in the channel estimation performance.	We see that DL-based approaches provide more robust performance and become resilient up to approximately 4 degrees angular mismatch between training and test data. We note here that more robustness may still be achieved if the training data will be enriched by adding more channel realizations with different angular information. The trade-off here is that if the number of channel realizations is high, the ability of the network to distinguish different input characteristics will be reduced due to the fact that mismatched channel matrices will become more similar, thus the network will yield the same, or least very similar, hybrid beamformer weights at the output. In our simulations, we have used $N=100$, $G=200$ channel realizations and add synthetic noise into those realizations to make the network more robust against mismatched data.	Since the main cause of the performance loss in Fig.~\ref{fig_SNR_AngleMismatch_Rate} is the channel estimation accuracy, we further investigate the channel estimation NMSE with respect to the angular mismatch in Fig.~\ref{fig_SNR_AngleMismatch_NMSE}. As it is seen, the angular mismatch deteriorates the channel estimation performance. These results support the robustness of the DL-based approaches obtained in Fig.~\ref{fig_SNR_AngleMismatch_Rate}.

	%	%%-----------------------------------------------------
	\begin{figure}[t!]
		\centering
		{\includegraphics[draft=false,width=\columnwidth,height=.7\columnwidth]{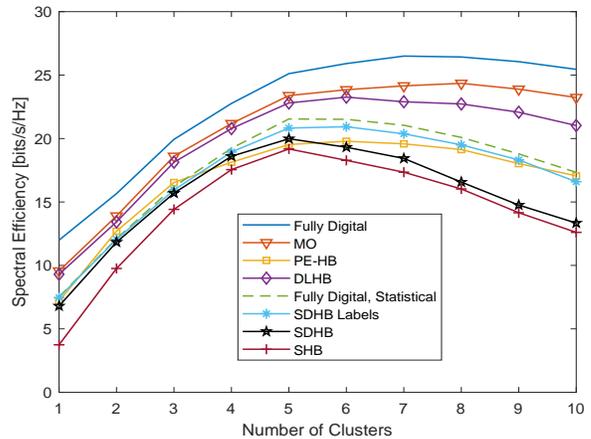} } 
		\caption{Spectral efficiency versus number of clusters $L$ when SNR$=0$ dB and SNR$_{\bar{\mathbf{N}}-\mathrm{TEST}}=10$ dB.   }
		\label{fig_SNR_NumberofPaths_Rate}
	\end{figure}
	
	%	%%-----------------------------------------------------

	%----------------------------------------------------------------------------------------------------------
	\begin{table*}[h]
		\caption{ Computation Time For Different Algorithms (In Seconds) }
		\label{tableComp}
		\centering
		\begin{tabular}{|c|c|c|c|c|c|c|c|c|c|c|}
			\hline
			\hspace{-3pt}	$N_\mathrm{T}$\hspace{-3pt}&\hspace{-3pt}Fully-Digital\hspace{-3pt}& \hspace{-3pt}Fully-Digital, Stat.\hspace{-3pt}&\hspace{-3pt}MO~\cite{hybridBFAltMin}&\hspace{-3pt} DLBH~\cite{elbirDL_COMML}\hspace{-3pt} &\hspace{-3pt} \textsf{CovNet}\hspace{-3pt}& \hspace{-3pt}\textsf{ChannelNet}\hspace{-3pt}& \hspace{-3pt}\textsf{BFNet}\hspace{-3pt}& SF-CNN~\cite{deepCNN_ChannelEstimation} \hspace{-3pt}&\hspace{-3pt} PE-HB~\cite{sohrabiNarrowband}&\hspace{-3pt} SHB~\cite{widebandHBWithoutInsFeedback} \hspace{-3pt}\\
			\hline
			\hline
			4& 0.0054&    0.0501&    0.6621&    0.0110&    0.0143&    0.0054&    0.0064&				0.0054&    0.0312&    0.0218 \\
			\hline
			8& 0.0057&    0.0535&    0.7247&    0.0117&    0.0145&    0.0061&    0.0073&    0.0072&    0.0358&    0.0292 \\
			\hline
			16&		0.0060&    0.0565&    1.6754&    0.0119&    0.0166&    0.0073&    0.0087&    0.0081&    0.0426&    0.0424\\
			\hline
			32&  0.0064&    0.0574&    1.6108&    0.0136&    0.0184&    0.0091&    0.0101&    0.0094&    0.0535&    0.0475\\
			\hline
			64&0.0079  &  0.0589 &   2.5603   & 0.0147 &   0.0211    &0.0108  &  0.0108   & 0.0103 &   0.0637  &  0.0481 \\
			\hline
			128&0.0083&    0.0612 &   4.4153 &   0.0165  &  0.0232   & 0.0118  &  0.0124  &  0.0123 &   0.0763 &   0.0590\\
			\hline
			
			\hline 
		\end{tabular}
	\end{table*}
	%----------------------------------------------------------------------------------------------------------

	Fig.~\ref{fig_SNR_NumberofPaths_Rate} illustrates the system rate performance of the algorithms with respect to the number of clusters $L$ when $N_\mathrm{RF}=4$. We see that the system rate increases when the spatial diversity of the channel is low (e.g., $L\leq 5$). When the diversity is high and the channel becomes less sparse, the performance of the algorithms deteriorates after $L> 5$ since the eigenvectors of $\mathbf{R}$ with respect to $N_\mathrm{RF}$ dominant eigenvalues do not represent the array response of the received clusters. We also observe that the increase in the number of clusters affects statistical approaches more significantly as compared to the non-statistical approaches. We have used, in this experiment, three $L$ values such as $L_\mathrm{TRAIN} = \{3,5,6\}$ when generating the training data so that robust performance can be obtained. Hence, the total length of the training dataset three times greater than the one given in Section~\ref{Training}. It can be seen from Fig.~\ref{fig_SNR_NumberofPaths_Rate} that SDHB has robust performance against different number of clusters even when $L> N_\mathrm{RF}$.

	%	%%-----------------------------------------------------
	\begin{figure}[t!]
		\centering
		{\includegraphics[draft=false,width=\columnwidth,height=.7\columnwidth]{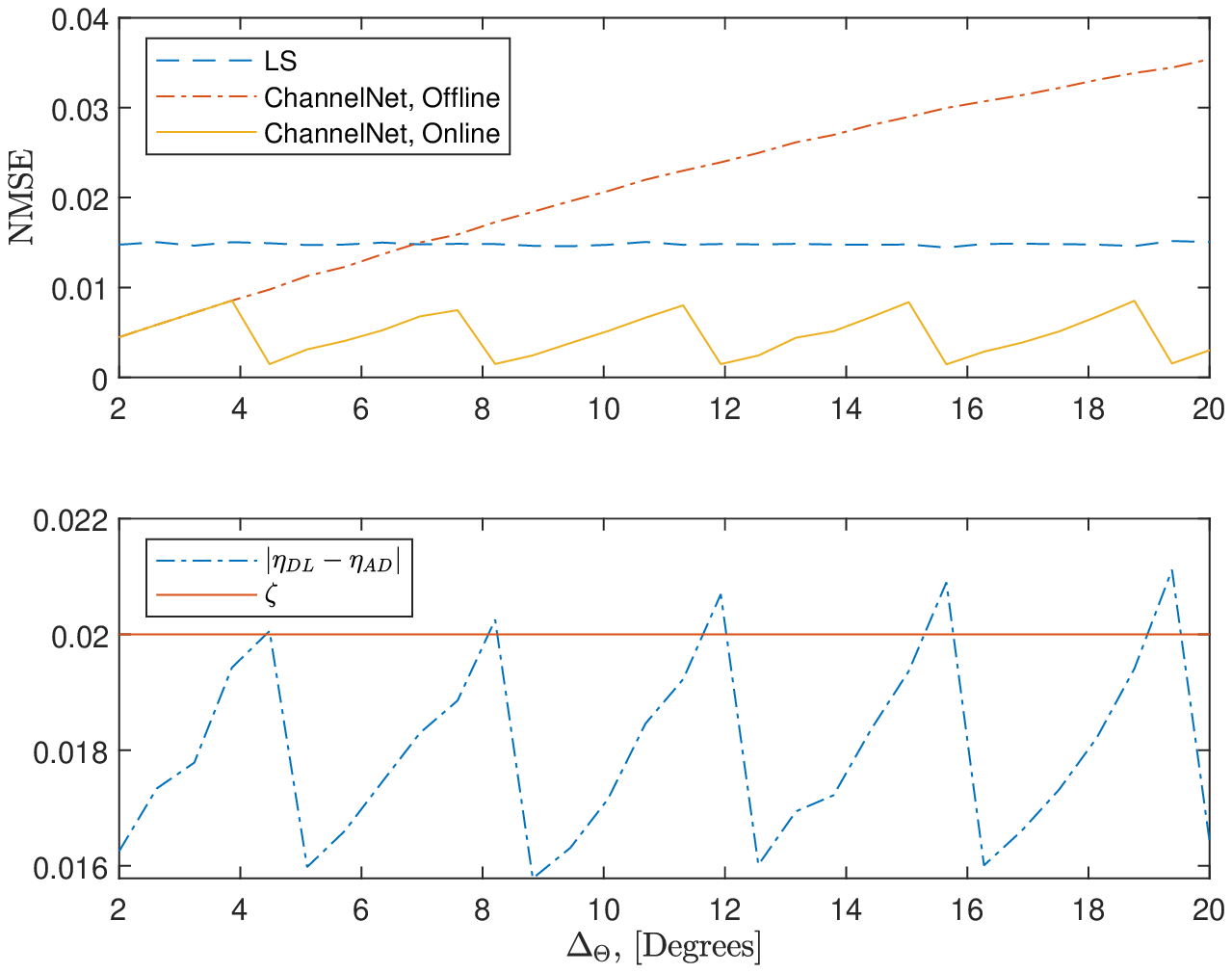} } 
		\caption{Online deployment for \textsf{ChannelNet} when SNR$=0$ dB and SNR$_{\bar{\mathbf{N}}-\mathrm{TEST}}=10$ dB.   }
		\label{fig_Online}
	\end{figure}
	
	%	%%-----------------------------------------------------

	We further investigate the computation time of the algorithms illustrated in Table~\ref{tableComp} for different number of BS antennas while the other parameters are fixed. We first compare the computation time of statistical and non-statistical fully-digital beamforming. We can see that statistical approach takes longer due to eigendecomposition of the CCM. Among all, the MO algorithm requires the longest time due to the involvement of the  optimization stage.  For a fair comparison, we can compare the computation time of \textsf{CovNet} and \textsf{BFNet} combined, with the other algorithms such as  MO, DLHB, PE-HB and SHB which do not involve channel estimation state. We can see that the proposed DL framework is the fastest algorithm among the all. Combining the complexity of all proposed deep networks \textsf{CovNet}, \textsf{ChannelNet} and \textsf{BFNet}, we obtain approximately the same complexity with comparison to  SHB which assumes perfect CSI. In contrast, compared to the MO algorithm, the proposed DL approach is at least 10 times faster (when $N_\mathrm{T}=4$, and 80 times faster if $N_\mathrm{T}=128$), which shows the potential of the proposed DL approach. 
	
	{\color{black} Above complexity analysis is valid with the assumption that the proposed DL model is trained beforehand. Therefore, we also discuss the offline training complexity. The convergence of the proposed neural network architecture takes about $2$ hours with the aforementioned training dataset and network settings. While this time duration is much larger than the computation time of the trained network, it is only performed once in the offline stage.

		Next, we examine the online prediction performance of \textsf{ChannelNet}, as illustrated in Fig.~\ref{fig_Online}.} In this experiment, the DL network is deployed online when there is $\Delta_\Theta = 2^\circ$ angle mismatch between the AOD/AOA angles of the training and the test data. Then we model the user motion such that $\Delta_\Theta$ changes from $2^\circ$ to $20^\circ$. We select $\zeta =0.02$, $G_\mathrm{Y}=200$ and use ADCE~\cite{angularDomainCE_feifei} algorithm to obtain the online labels as described in Algorithm~\ref{alg:OnlineDeploy}. We can see that the performance of the offline network gets poorer as $\Delta_\Theta$ increases since its performance degrades due to the mismatch of the new incoming data. For the online deployment case, the network is updated when the update rule triggers so that the network adapts the environment and its performance gets better when updated. It is shown that network requires to be re-trained for approximately every $4^\circ$ mismatch, similar to the observations in Fig.~\ref{fig_SNR_AngleMismatch_Rate}. {\color{black}To account for the time complexity, the online update (i.e., training with the online dataset) of the network only takes about $0.6$ s while offline training overhead is about $2$ hours. This is mainly because of the use of small online dataset. Nevertheless, such small dataset provides very good NMSE performance. In addition, the computation time for employing the network for channel estimation is similar to the results given in Table~\ref{tableComp}. As a result, the online training not only improves the estimation performance, but also accelerates the network training to adapt to the new environment.}

	\section{Summary}
	\label{sec:Conc}
	We introduced a DL framework for hybrid beamforming and channel estimation for mm-Wave massive MIMO systems without instantaneous CSI feedback. We designed three CNNs, one of which, \textsf{CovNet} is used at the BS to design the precoders by using the channel covariance matrix. Two CNNs, \textsf{ChannelNet} and \textsf{BFNet} are placed at the receiver for channel estimation and combiner design, respectively. 
	{\color{black} We have examined the online deployment of the proposed DL-based channel estimation scheme and shown that  the proposed approach can adapt itself to the propagation environment and update its parameters accordingly. We have also shown that the proposed scheme can work properly up to $4$ degrees angular mismatch and do not need to be retrained.} In another challenging experiment where we present the performance against the number of clusters, we have shown that the proposed approach has robust performance even when the number of clusters differs in the training and the test data.	{\color{black}In addition, we have shown that the proposed approach can perform at least $10$ times faster than the optimization based approaches in online deployment. It is also worthwhile to mention that the offline training takes much longer than online computation time, however this process is only performed once. Moreover, the proposed online training scheme reduces the training overhead significantly without the requirement of re-training the DL model from scratch.    }

	\bibliographystyle{IEEEtran}
	\footnotesize{\bibliography{IEEEabrv,references_050_journal}}
	%	\balance
	\begin{IEEEbiography}[{\includegraphics[width=1in,height=1.25in,clip,keepaspectratio]{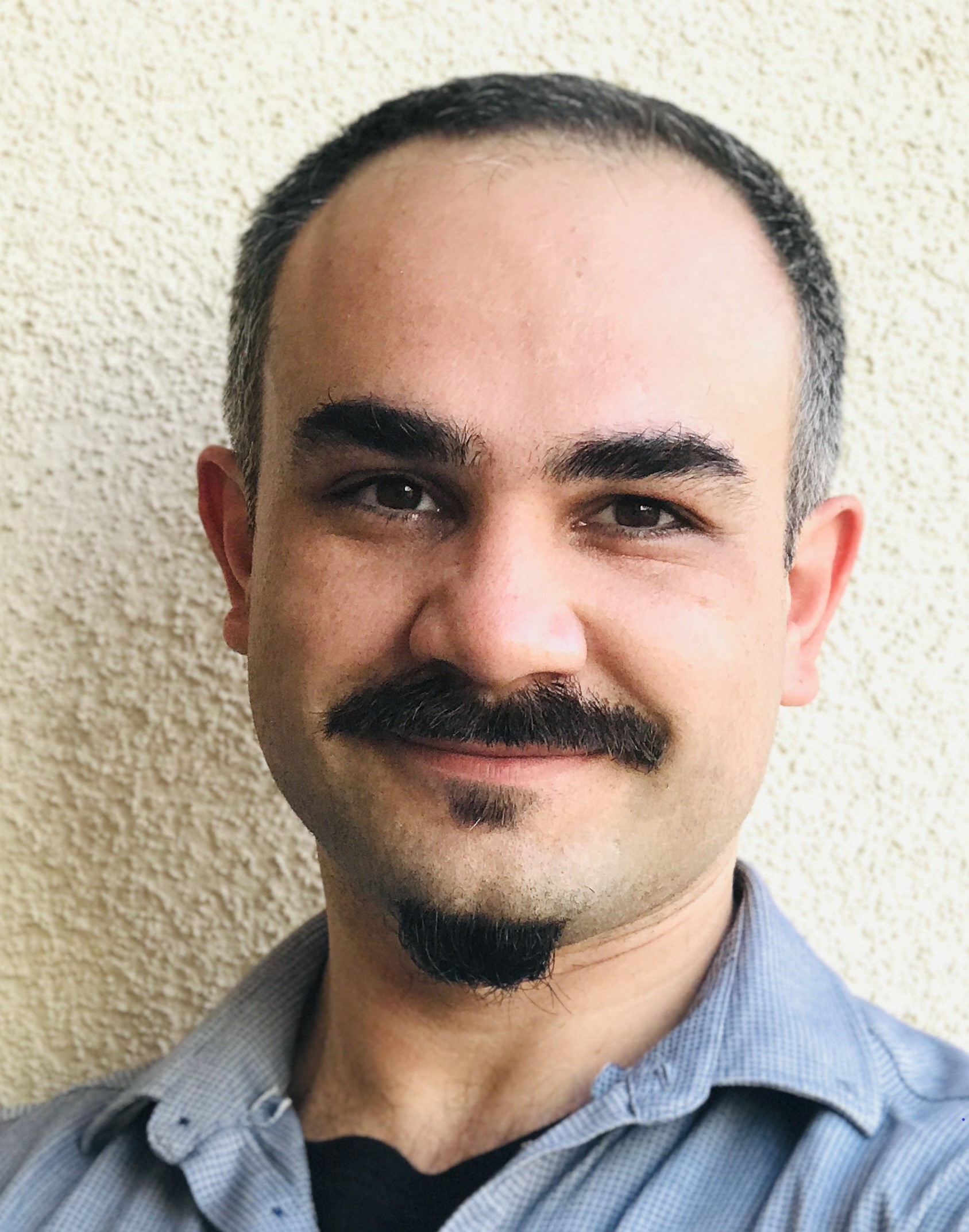}}]{Ahmet M. Elbir}	(Senior Member, IEEE) received		the B.S. degree (with Hons.) in electrical engineering from Firat University in 2009, and the Ph.D. 		degree in electrical engineering from Middle East		Technical University (METU) in 2016. He is currently a Visiting Postdoctoral Researcher with Koc	University, and a Research Fellow with Duzce University.  His research interests 		include array signal processing, sparsity-driven convex optimization, signal processing for communications, and deep learning for array signal processing. 		He was a recipient of the 2016 METU Best Ph.D.	Thesis Award for his Doctoral Studies. He has been serving as an Associate	Editor for IEEE ACCESS and Frontiers in Communications and Networks.
	\end{IEEEbiography}

\end{document}